\documentclass[english,11pt,aps,pra,onecolumn,tightenlines,groupedaddress,superscriptaddress,notitlepage,floatfix,fleqn]{revtex4-1}
\pdfoutput=1
\usepackage[utf8]{inputenc}
\usepackage[T1]{fontenc}
\usepackage{amsmath}
\usepackage{amssymb}
\usepackage{amsfonts}
\usepackage{bm}
\usepackage{bbm}
\usepackage{epsfig}
\usepackage{times}
\usepackage{amsthm}
\usepackage{graphics}

\usepackage[usenames,dvipsnames]{color}
\usepackage{tikz}
\definecolor{grey}{rgb}{.35,.35,.35}
\definecolor{dblue}{rgb}{0,0.1,.6}
\definecolor{dgreen}{rgb}{0,.6,0.1}

\newcommand{\tentative}[1] {}

\usepackage[colorlinks=true,citecolor=dblue,linkcolor=dblue,urlcolor=dblue]{hyperref}
\usepackage[all]{hypcap}

\newcommand{\id}{\mathbbm{1}}
\newcommand{\bra}{\langle}
\newcommand{\ket}{\rangle}
\newcommand{\Tr}{\operatorname{Tr}}
\newcommand{\Span}{\operatorname{span}}
\newcommand{\Rank}{\operatorname{rank}}
\newcommand{\dist}{\operatorname{dist}}
\renewcommand{\vec}[1]{{\boldsymbol{#1}}}

\newcommand{\hH}{\hat{H}}
\newcommand{\hS}{\hat{S}}
\newcommand{\hA}{\hat{A}}
\newcommand{\hB}{\hat{B}}
\newcommand{\hD}{\hat{D}}
\newcommand{\hP}{\hat{P}}

\newcommand{\hZ}{\hat{Z}}
\newcommand{\hT}{\hat{t}}
\newcommand{\hU}{\hat{U}}
\newcommand{\hV}{\hat{V}}
\newcommand{\dm}{{\hat{\rho}}}

\newcommand{\s}{\sigma}
\newcommand{\vs}{\vec{\sigma}}
\newcommand{\NN}{\mathbb{N}}
\newcommand{\ZZ}{\mathbb{Z}}
\newcommand{\RR}{\mathbb{R}}
\newcommand{\CC}{\mathbb{C}}

\newcommand{\hvS}{\hat{\vec{S}}}

\newcommand{\mc}[1]{\mathcal{#1}}

\renewcommand{\H}{\mc{H}}
\newcommand{\A}{\mc{A}}
\newcommand{\B}{\mc{B}}
\newcommand{\V}{\mc{V}}
\newcommand{\E}{\mc{E}}
\newcommand{\G}{\mc{G}}
\newcommand{\M}{\mc{M}}
\newcommand{\In}{\operatorname{in}}
\newcommand{\Out}{\operatorname{out}}

\newcommand{\inj}{\text{inj}}
\newcommand{\pMPS}{\operatorname{pMPS}}
\newcommand{\pMPSn}{\operatorname{pMPS_n}}
\newcommand{\ipMPS}{\operatorname{ipMPS}}
\newcommand{\oMPS}{\operatorname{oMPS}}
\newcommand{\tiMPS}{\operatorname{tiMPS}}
\newcommand{\TTNS}{\operatorname{TTNS}}
\newcommand{\MERA}{\operatorname{MERA}}
\newcommand{\PEPS}{\operatorname{PEPS}}
\newcommand{\R}{\operatorname{R}}
\newcommand{\Gr}{\operatorname{Gr}}
\newcommand{\St}{\operatorname{St}}
\newcommand{\U}{\operatorname{U}}
\newcommand{\Q}{\operatorname{Q}}
\newcommand{\End}{\operatorname{End}}
\newcommand{\GL}{\operatorname{GL}}
\newcommand{\SL}{\operatorname{SL}}

\newcommand{\RQ}{\text{RQ}}
\newcommand{\AKLT}{\text{AKLT}}
\newcommand{\veps}{\varepsilon}
\newcommand{\vphi}{\varphi}

\newcommand{\bigtimes}{\mathop{\tikz [x=1.2ex,y=1.6ex,line width=.2ex] \draw (0,0) -- (1,1) (0,1) -- (1,0);}}

\renewcommand{\Bmatrix}[1]{\begin{bmatrix}#1\end{bmatrix}}

\newcommand{\Emph}[1]{\textbf{\emph{#1}}}

\newtheorem{prop}{Proposition}
\newtheorem*{theorem}{Theorem}

\newcommand{\duke}  {Department of Physics, Duke University, Durham, North Carolina 27708, USA}
\newcommand{\dukem} {Departments of Mathematics, Chemistry, and Physics, Duke University, Durham, North Carolina 27708, USA}
\newcommand{\tum}   {Department of Mathematics, Technical University of Munich, D 85748 Garching, Germany}

\begin{document}

\title{On the Closedness and Geometry of Tensor Network State Sets}
\author{Thomas Barthel}
\affiliation{\duke}
\author{Jianfeng Lu}
\affiliation{\dukem}
\author{Gero Friesecke}
\affiliation{\tum}

\begin{abstract}
Tensor network states (TNS) are a powerful approach for the study of strongly correlated quantum matter. The curse of dimensionality is addressed by parametrizing the many-body state in terms of a network of partially contracted tensors. These tensors form a substantially reduced set of effective degrees of freedom. In practical algorithms, functionals like energy expectation values or overlaps are optimized over certain sets of TNS. Concerning algorithmic stability, it is important whether the considered sets are closed because, otherwise, the algorithms may approach a boundary point that is outside the TNS set and tensor elements diverge. We discuss the closedness and geometries of TNS sets, and we propose regularizations for optimization problems on non-closed TNS sets. We show that sets of matrix product states (MPS) with open boundary conditions, tree tensor network states (TTNS), and the multiscale entanglement renormalization ansatz (MERA) are always closed, whereas sets of translation-invariant MPS with periodic boundary conditions (PBC), heterogeneous MPS with PBC, and projected entangled-pair states (PEPS) are generally not closed. The latter is done using explicit examples like the W state, states that we call two-domain states, and fine-grained versions thereof.
\end{abstract}

\date{April 15, 2022}

\maketitle

\renewcommand{\baselinestretch}{0.85}\normalsize
\tableofcontents
\renewcommand{\baselinestretch}{1}\normalsize
\section{Introduction}
We consider quantum many-body systems on a lattice with $N$ sites \footnote{We typically think of condensed matter systems defined on, say, the $D$-dimensional hypercubic lattice $\{1,2,\dotsc,L\}^{\times D}$ with $N=L^D$ sites. However, tensor network methods are applicable to a wide variety of problems, and a \emph{lattice site} would for example correspond to an orbital in quantum chemistry applications.} and $d$-dimensional single-site Hilbert spaces $\H_1\simeq\CC^{d}$ with orthonormal basis $\{|\s\ket\,|\,\s=1,\dotsc,d\}$. The total Hilbert space is the $N$-fold tensor product $\H=\H_1^{\otimes N}$ and we use the notations $|\vs\ket:=|\s_1,\s_2,\dotsc,\s_N\ket:=|\s_1\ket\otimes|\s_2\ket\otimes\dotsb\otimes|\s_N\ket$ for its orthonormal basis of tensor product states. To improve the readability, we usually fix $d$ for all sites, and only use site-dependent dimensions $d_i$ when necessary.

The exponential growth of the total Hilbert space dimension in $N$ is sometimes called the curse of dimensionality and limits exact treatments of most quantum many-body systems to very small system sizes $N$.
Tensor network states (TNS) $|\Psi\ket$ are (approximate) representations of states $|\psi_0\ket\in\H$, where expansion coefficients $\bra\vs|\Psi\ket$ take the form of a tensor network, i.e., a network of partially contracted tensors. The only non-contracted (``open'') indices label the local basis states $|\sigma_i\ket$. The other indices are referred to as virtual or bond indices. Each bond index occurs at exactly two individual tensors, and one obtains $\bra\vs|\Psi\ket$ by summing over all bond indices. TNS methods are a powerful approach for the investigation of strongly-correlated quantum many-body systems, employed mostly in condensed matter physics and quantum chemistry. The tensor network forms a graph, where the tensors are associated to nodes of the graph, and the bond indices are associated with the edges of the graph \cite{Orus2014-349}. When the network structure is well-aligned with the entanglement structure of $|\psi_0\ket$, good TNS approximations of $|\psi_0\ket$ can be achieved with relatively small bond dimensions (small numbers of effective degrees of freedom). For this, it is desirable that strongly entangled sites have a small graph distance. For condensed matter problems, a common choice is to arrange the graph according to spatial distances in the physical system. For quantum chemistry problems, the choice is less obvious and may also be determined numerically \cite{Legeza2003-68,Chan2002-116,Rissler2006-323,Krumnow2016-117}.
For one-dimensional systems, one can often reach machine precision accuracy \cite{White1992-11,White1993-10,Schollwoeck2005,Hastings2007-08,Verstraete2005-5,Landau2015-11,Barthel2017_08}. In contrast to quantum Monte Carlo methods, TNS simulations are not hampered by the negative sign problem \cite{Loh1990-41,Chandrasekharan1999-83,Troyer2005} for frustrated quantum magnets and fermionic systems \cite{Barthel2009-80,Corboz2009-80,Pineda2009_05,Kraus2009_04,Corboz2009_04}.

In this paper, we address the question of whether various sets of TNS are closed or not closed in the sense of topology \footnote{Recall that a subset of a Hilbert space is closed if, for every sequence of states \unexpanded{$\Psi_n$} in the subset which converges to some limit state \unexpanded{$\Psi^*$}, the latter is contained in the subset. Here, convergence means strong convergence, i.e., \unexpanded{$\lim_{n\to\infty}\|\Psi_n-\Psi^*\|=0$}. With the exception of Sec.~\ref{sec:infiniteTNS}, we consider finite-dimensional Hilbert spaces such that the notions of strong and weak convergence are equivalent.}. The answer has fundamental implications for TNS optimization algorithms. Typical goals are the minimization of energy expectation values $\bra\Psi|\hH|\Psi\ket/\|\Psi\|^2$ to compute ground states, the minimization of energy expectation values under orthogonality constraints $\bra \psi_i|\Psi\ket=0$ to compute excited energy eigenstates, and the maximization of overlaps $|\bra\psi_0|\Psi\ket|$, e.g., in order to find efficient approximations of states $|\psi_0\ket=\hat{X}|\Phi\ket$ obtained by acting with an operator $\hat{X}$ on a TNS $|\Phi\ket$. The latter task arises, for example, in real-time evolution algorithms and in imaginary-time evolution or power methods for the computation of ground states.
The optimization over a non-closed set of TNS may be driven to a boundary point outside the TNS set, i.e., optimizers may not exist. In such a case, tensor elements of the TNS representation diverge, and the algorithm becomes unstable.

The simplest type of TNS are matrix product states (MPS) \cite{Baxter1968-9,Accardi1981,Fannes1992-144,White1992-11,Rommer1997,PerezGarcia2007-7,Schollwoeck2011-326}. They are at the heart of the famous density-matrix renormalization group (DMRG) algorithm \cite{White1992-11,Rommer1997,Schollwoeck2011-326}. An MPS is characterized by assigning an order-3 tensor $A_i\in \CC^{d\times m\times m'}$ to each lattice site $i$. We will often interpret such a tensor as a collection of $m\times m'$ matrices $A^{1},\dotsc, A^{d}$, and the dimensions $m$ and $m'$ of the virtual indices are called bond dimensions. We discuss three different classes of MPS: those with open boundary conditions (OBC), those with periodic boundary conditions (PBC), and those with PBC and translation invariance. Further classes of TNS that we address are tree tensor network states (TTNS) \cite{Shi2006-74,Murg2010-82}, the multiscale entanglement renormalization ansatz (MERA) \cite{Vidal-2005-12,Vidal2006}, and projected entangled-pair states (PEPS) in $D>1$ spatial dimensions \cite{Niggemann1997-104,Nishino2000-575,Martin-Delgado2001-64,Verstraete2004-7,Verstraete2006-96}.
Some alternative terms used in the mathematical literature are \emph{tensor train} for an MPS with OBC \cite{Oseledets2011-33}, \emph{tensor ring} for an MPS with PBC, and \emph{hierarchical Tucker format} for a TTNS \cite{Hackbusch2009-15,Grasedyck2010-31}.

The main part of the paper addresses TNS for finite system sizes $N$. Using the TNS gauge freedom, it is easy to see that OBC-MPS form closed sets, which are intimately related to direct products of Grassmann manifolds (Sec.~\ref{sec:oMPS}). This generalizes immediately to TTNS (Sec.~\ref{sec:TTNS}). Similarly, MERA states form closed sets due to the isometric property of their tensors (Sec.~\ref{sec:MERA}).
Sets of normalized PBC-MPS with fixed bond dimensions are generally non-closed. We present proofs for PBC-MPS with and without translation-invariant tensors (Secs.~\ref{sec:tiMPS} and \ref{sec:pMPS}). For heterogeneous PBC-MPS, we rephrase and extend results of Ref.~\cite{Landsberg2012-12} for the case $d=m^2$, such that the practically relevant scenarios with $d\ll m$ are now covered.
These results generalize to PEPS in $D>1$ dimensions, which generally form non-closed sets if the network contains loops (Sec.~\ref{sec:PEPS}). In Sec.~\ref{sec:infiniteTNS}, we shortly comment on tensor networks for infinite systems \cite{Vidal2007-98,Orus2008-78,McCulloch2008_04,Zauner2018-97,Jordan2008-101,Orus2009_05,Evenbly2009-79,Montangero2008-10}.
We close in Sec.~\ref{sec:discuss} with a summarizing theorem, examples of optimization problems on non-closed TNS sets with and without optimizers, a geometric intuition for the occurrence of non-included boundary points, implications for TNS algorithms, and suggestions for corresponding regularizations.

It has recently been pointed out that, given the tensor of an infinite PEPS (iPEPS), it is generally undecidable whether the state has a nonzero norm or a certain symmetry \cite{Scarpa2020-125}. Our results imply that iPEPS optimization problems, even if they are driven to a normalizable state, may be driven to a non-included point on the boundary of the iPEPS set. Furthermore, the infinite system size and translation invariance are not the decisive factors in this respect; the same problem can occur for heterogeneous PEPS on finite systems.

\section{Matrix product states with OBC}\label{sec:oMPS}
The set $\oMPS(N,\{m_i\},d)$ of normalized MPS with OBC contains all states of the form
\begin{equation}\label{eq:oMPS}\textstyle
	|\Psi\ket=\sum_\vs A_1^{\s_1}A_2^{\s_2}\dotsb A_N^{\s_N}|\vs\ket\quad\text{with}\quad
	\|\Psi\|:=\sqrt{\bra\Psi|\Psi\ket}=1,
\end{equation}
characterized by $N$ order-3 tensors $A_i\in \CC^{d\times m_{i-1}\times m_{i}}$. The first and last bond dimensions need to be $m_0=m_N=1$ in order for the matrix product in Eq.~\eqref{eq:oMPS} to give a scalar coefficient.

\begin{prop}\label{prop:oMPS}
The set $\oMPS(N,\{m_i\},d)$ is closed and coincides with the set of normalized states in $\H$ with Schmidt ranks $\tilde{m}_i\leq m_i$ for all bipartitions of the system into blocks of sites $[1,i]$ and $[i+1,N]$.
\end{prop}
\Emph{Proof:}
\textbf{(a)} Call $\H_\A=\H_1^{\otimes i}$ the Hilbert space for the block of sites $\A=[1,i]\equiv\{1,\dotsc,i\}$ and $\H_\B=\H_1^{\otimes (N-i)}$ the Hilbert space of the remaining sites. The Schmidt decomposition \cite{Nielsen2000} of a state $|\psi\ket\in\H=\H_\A\otimes\H_\B$ brings it into the form $|\psi\ket=\sum_{k=1}^{\tilde{m}}\lambda_k |k\ket_\A\otimes|k\ket_\B$ with Schmidt coefficients $\lambda_k>0$ and sets of orthonormal states $\{|k\ket_\A\in\H_\A\}$ and $\{|k\ket_\B\in\H_\B\}$. We call $\tilde{m}$ the Schmidt rank of $|\psi\ket$, and we refer to $\Span\{|k\ket_\A\}$ and $\Span\{|k\ket_\B\}$ as the (uniquely determined) Schmidt spaces.
The set $\R(N,\{m_i\},d)\subseteq\H$ of normalized states with Schmidt ranks $\tilde{m}_i\leq m_i$ for all such bipartitions is a closed set: For each bipartition, regard $\bra\vs|\psi\ket=:\psi_i^{(\s_1,\dotsc,\s_i),(\s_{i+1},\dotsc,\s_N)}$ as the elements of the $d^i\times d^{N-i}$ matrix $\psi_i$. Now, let $\vec{M}(|\psi\ket)$ denote the complex vector formed by the list of all $(m_i+1)\times (m_i+1)$ minors of the matrices $\psi_i$ for $i\in[1,N]$. Recall that a matrix has rank $m$ if and only if some $m\times m$ minor is nonzero and every $(m+1)\times(m+1)$ minor vanishes. $\vec{M}(|\psi\ket)$ is the zero-vector $\vec{0}$ if and only if $|\psi\ket$ has Schmidt ranks $\tilde{m}_i\leq m_i$ $\forall_i$. So, $\R(N,\{m_i\},d)$ is the preimage of the closed set $\{\vec{0}\}$ with respect to the continuous function $\vec{M}$ and is hence itself a closed set.

\textbf{(b)} Every MPS \eqref{eq:oMPS} has Schmidt ranks $\tilde{m}_i\leq m_i$ because we can write it as a sum of $m_i$ tensor products: $|\Psi\ket=\sum_{n=1}^{m_i}|n\ket_\A\otimes |n\ket_\B$ with
\begin{subequations}\label{eq:oMPSsplit}
\begin{alignat}{5}
	&|n\ket_\A&&=\textstyle\sum_{\s_1,\dotsc,\s_i}     &&[A_{1}^{\s_{1}}    \dotsb A_{i}^{\s_i}]_{1,n}\,&&|\s_{1},  \dotsc,\s_i\ket\quad\text{and}\\\label{eq:oMPS_nB}
	&|n\ket_\B&&=\textstyle\sum_{\s_{i+1},\dotsc,\s_N} &&[A_{i+1}^{\s_{i+1}}\dotsb A_N^{\s_N}  ]_{n,1}\,&&|\s_{i+1},\dotsc,\s_N\ket.
\end{alignat}
\end{subequations}
From this, one obtains a Schmidt decomposition with $\tilde{m}_i\leq m_i$ components, e.g., by diagonalization of the reduced density matrix $\dm_\A=\Tr_\B|\Psi\ket\bra\Psi|$ for subsystem $\A=[1,i]$.

\begin{figure*}[t]
\label{fig:oMPS}
\includegraphics[width=\textwidth]{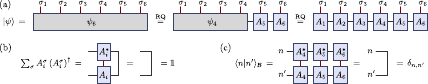}
\caption{(a) Decomposition of a state $|\psi\ket$ into an MPS with OBC \eqref{eq:oMPS} by a sequence of five RQ decompositions \eqref{eq:oMPS_fromPsi} for a system of $N=6$ sites. (b) The resulting tensors $A_i$ can be interpreted as isometries. (c) This isometric property guarantees that the resulting block states $|n\ket_\B$ as defined in Eq.~\eqref{eq:oMPS_nB} are orthonormal and that bond dimensions agree with the Schmidt ranks of the state $|\psi\ket$.}
\end{figure*}

\textbf{(c)} Every state $|\psi\ket=\sum_\vs \psi_N^{\s_1,\dotsc,\s_N}|\vs\ket$ with Schmidt ranks $\tilde{m}_i$ can be written as an MPS \eqref{eq:oMPS} with bond dimensions $m_i=\tilde{m}_i$: The MPS tensors $A_i\in\CC^{d\times m_{i-1}\times m_i}$ can be determined by a sequence of RQ decompositions, e.g., starting at the right end of the chain as illustrated in Fig.~\ref{fig:oMPS}. 
\begin{equation}\label{eq:oMPS_fromPsi}
	\psi_N^{\s_1,\dotsc,\s_{N-1},\s_N}\stackrel{\RQ}{=:}\psi_{N-1}^{\s_1,\dotsc,\s_{N-1}} A_N^{\s_N} \quad\text{and}\quad
	\psi_i^{\s_1,\dotsc,\s_{i-1},\s_i}\stackrel{\RQ}{=:}\psi_{i-1}^{\s_1,\dotsc,\s_{i-1}} A_i^{\s_i}\quad\text{for}\quad
	i=N-1,\dotsc,1.
\end{equation}
In each step, $\psi_i$ is treated as a $d^{i-1}\times (d m_i)$ matrix to do the RQ factorization, and $m_N=1$. We are using the reduced (or ``thin'') RQ decomposition \cite{Golub1996} such that bond dimension $m_{i-1}$ is the rank of $\psi_i$ in matrix form. The isometric property of the ``Q'' matrix ($QQ^\dag=\id$) translates into the property $\sum_\s A^\s_i A^{\s\dag}_i=\id$ and implies that $\left\{|n\ket_\B\,|\, n=1,\dots,m_i\right\}$, with $|n\ket_\B$ as defined in Eq.~\eqref{eq:oMPS_nB}, is an orthonormal basis for the Schmidt space of sites $\B=[i+1,N]$. Hence, the obtained bond dimensions $m_i$ agree with the Schmidt ranks $\tilde{m}_i$ of $|\psi\ket$. In fact, the procedure could similarly be done using singular value decompositions (SVD), which, in every step, would correspond to a Schmidt decomposition of the state.

In conclusion, the set  $\R(N,\{m_i\},d)$ of Schmidt-rank bounded and normalized states is closed and $\oMPS(N,\{m_i\},d)=\R(N,\{m_i\},d)$. \qed

There is an alternative, slightly quicker way of seeing that $\oMPS(N,\{m_i\},d)$ is a closed set: Any OBC-MPS \eqref{eq:oMPS} is invariant under \emph{gauge transformations}
\begin{equation}\label{eq:gaugeTrafo}
	\left(A_i^\s, \ A_{i+1}^{\s'}\right) \ \mapsto \ \left(A_i^\s Z^{-1}_i, \ Z_i A_{i+1}^{\s'}\right) \ \forall_{\s,\s'} \quad\text{with} \quad i\in[1,N-1]
\end{equation}
and invertible $m_i\times m_i$ matrices $Z_i$. This can be used to bring the matrix product into a so-called \emph{orthonormal form}. With a sequence of reduced RQ decompositions, starting at the right end of the chain,
\begin{equation}\label{eq:oMPS_RQ}
	A_N^{\s_N}\stackrel{\RQ}{=:}R_{N-1} B_N^{\s_N} \quad\text{and}\quad
	A_{i}^{\s_{i}}R_{i}\stackrel{\RQ}{=:}R_{i-1} B_{i}^{\s_{i}}
	\quad\text{for}\quad
	i=N-1,N-2,\dotsc,1
\end{equation}
where $R_{1}=\|\Psi\|=1$ and $Z_i\hat{=}R_i^{-1}$ in Eq.~\eqref{eq:gaugeTrafo}, one arrives at a so-called right-canonical form of the MPS
\begin{equation}\label{eq:oMPSon}\textstyle
	|\Psi\ket=\sum_\vs B_1^{\s_1}B_2^{\s_2}\dotsb B_N^{\s_N}|\vs\ket \quad\text{with}\quad
	\sum_{\s=1}^{d} B_i^\s B_i^{\s\dag}=\id.
\end{equation}
The reduced dimensions $\tilde{m}_i\leq m_i$ of tensor $B_i\in \CC^{d\times \tilde{m}_{i-1}\times \tilde{m}_{i}}$ obey $\tilde{m}_i\leq d \tilde{m}_{i\pm 1}$.
Tensor $B_i$ can be interpreted as an element of the Stiefel manifold $\St(\tilde{m}_{i-1},d \tilde{m}_i)$ with $\St(k,n)\equiv\{B\in \CC^{k\times n}\,|\,BB^\dag=\id_k\}\simeq\U(n)/\U(n-k)$ and $k\leq n$. One can remove a remaining unitary gauge freedom, which makes $B_i$ an element of the Grassmann manifold $\Gr(\tilde{m}_{i-1},d \tilde{m}_i)$ with $\Gr(k,n)\equiv\St(k,n)/\U(k)$. Due to the isometry constraint $\sum_{\s} B_i^\s B_i^{\s\dag}=\id$, the tensor elements are bounded by $\left|[B^\s_i]_{n,n'}\right|\leq \sqrt{\tilde{m}_{i-1}}$ $\forall_{i,\s,n,n'}$. So, the set $\oMPS(N,\{m_i\},d)$ is the image of the bounded and closed set
\begin{equation}\label{eq:oMPS-Gr}\textstyle
	\bigcup_{1\leq\tilde{m}_i\leq m_i} \U(1)\times\Gr(\tilde{m}_0,d \tilde{m}_1)\times \Gr(\tilde{m}_1,d \tilde{m}_2) \times \dotsb\times \Gr(\tilde{m}_{N-1},d \tilde{m}_N)
\end{equation}
with respect to the continuous map \eqref{eq:oMPSon}. It is hence itself a closed set. With the real dimension $\dim_\RR \Gr(k,n)=2\dim_\CC \Gr(k,n)=2k(n-k)$ of the Grassmannian, the real dimensions of the manifolds in the union \eqref{eq:oMPS-Gr} are
\begin{equation}\textstyle
	 2\sum_{i=1}^{N} d \tilde{m}_i \tilde{m}_{i-1} - 2\sum_{i=1}^{N-1} \tilde{m}_i^2 - 1\quad
	\text{with}\quad \tilde{m}_i\leq d \tilde{m}_{i\pm 1}.
\end{equation}

The result that the Schmidt-rank-bounded set $\R(N,\{m_i\},d)$ is closed is certainly known to experts. It follows, for instance, from Theorem~3 in Ref.~\cite{Holtz2012-120}, stating that this set is a finite union of embedded submanifolds, or from a result of Hackbusch on hierarchical tensor spaces discussed at the end of Sec.~\ref{sec:TTNS}. The simple proofs given above and the characterization as the set $\oMPS(N,\{m_i\},d)$ are to our knowledge not contained in the previous literature.

\section{Tree tensor network states}\label{sec:TTNS}
A TTNS $|\Psi\ket$ is a TNS defined for a connected graph with vertices $i=1,\dots,N$ and a set $\E$ of edges without loops, i.e., a graph that decomposes into two disconnected subgraphs if we remove any edge $e\in\E$. We will denote the two components of the corresponding bipartition of the system $\A_e$ and $\B_e$. The edges $e$ are associated with bond vector spaces of dimensions $m_e$ and each vertex $i$ is associated with a site Hilbert-space of dimension $d_i$. So the network structure is characterized by $\G=(\E,\{m_e\},\{d_i\})$.
Here, we explicitly denote vertex-dependent $d_i$, because it is, for example, very common to have TTNS with internal tensors that carry no physical index $\sigma_i$, i.e., $d_i=1$ for such a vertex.

For a given network $\G$, a TTNS $|\Psi\ket$ is characterized by assigning a tensor $A_i\in\CC^{d_i\times \prod_{e\in\partial i}m_e}$ to each vertex $i$ and contracting the tensors $A_i$, i.e., summing over all their bond indices $n_e=1,\dots,m_e$. Here, $\partial i\subseteq\E$ is the set of edges connected to vertex $i$. The set of such normalized TTNS is denoted by $\TTNS(\G)$.

\begin{figure*}[t]
\label{fig:TTNS}
\includegraphics[width=\textwidth]{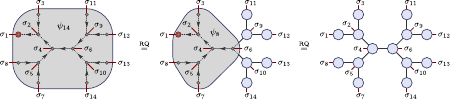}
\caption{Decomposition of a state $|\psi\ket$ into a TTNS for a system with $N=14$ sites as discussed in Sec.~\ref{sec:TTNS}. Site $i_0=1$ is chosen as the root and edges directed accordingly. One then splits off TTNS tensors sequentially using RQ decompositions, starting from the site with the largest distance from $i_0$. The resulting bond dimensions agree with the Schmidt ranks of the state $|\psi\ket$.}
\end{figure*}

\begin{prop}\label{prop:TTNS}
The set $\TTNS(\G)$ for a network $\G=(\E,\{m_e\},\{d_i\})$ is closed and coincides with the set of normalized states in $\H$ with Schmidt ranks $\tilde{m}_e\leq m_e$ for all bipartitions of the system into components $\A_e$ and $\B_e$.
\end{prop}
\Emph{Proof:} This is a straightforward generalization of the OBC-MPS case. With the same argument as under point (a) of Sec.~\ref{sec:oMPS}, the set $\R(\G)$ of normalized states in $\H$ with Schmidt ranks $\tilde{m}_e\leq m_e$ for the bipartitions of the system into components $\A_e$ and $\B_e$ is a closed set. With the same argument as under point (b) of Sec.~\ref{sec:oMPS}, $\TTNS(\G)\subseteq \R(\G)$. Also, $\R(\G)\subseteq \TTNS(\G)$ according to the following scheme for writing $|\psi\ket=\sum_\vs \psi^{\s_1,\dotsc,\s_N}|\vs\ket\,\in\R(\G)$ as a TTNS with the desired graph structure and bond dimensions $\{m_e\}$:

Designate a leaf $i_0$ of the tree as its root. Assign directions to all edges $e\in \E$, such that $e$ points from a vertex $i$ to a vertex $i'$ if we have $\dist(i',i_0)<\dist(i,i_0)$ for their graph distances from the root. $\A_e$ is then chosen as the component that contains $i'$ (and $i_0$) and its complement $\B_e$ contains $i$.
As in Eq.~\eqref{eq:oMPS_fromPsi}, we will decompose $\psi^{\s_1,\dotsc,\s_N}$ into a tree tensor network by a sequence of reduced RQ decompositions, splitting off one single-site tensor $A^{\s_i}_i$ at a time. In each step, we have a tensor $\psi^{\s_{i_1},\dotsc,\s_{i_k}}_{n_{e_1},\dotsc,n_{e_\ell}}$ that carries $k$  physical indices and $\ell$ bond indices, where,  initially, $k=N$ and $\ell=0$. We then select a vertex $i$ from $\{i_1,\dotsc,i_k\}$ that has maximal graph distance from the root $i_0$, and obtain tensor $A_i^{\s_i}$ from $\psi$ in an RQ decomposition as in Eq.~\eqref{eq:oMPS_fromPsi} or, equivalently, using an SVD. This gives a new tensor $\psi$ with one less physical index and a changed set of bond indices. In particular, it will now carry the bond index for bond $e$ that starts at vertex $i$. If we use an SVD, this decomposition is in fact a Schmidt decomposition of the global state and the resulting bond dimension agrees with the Schmidt rank $\tilde{m}_e$ for the bipartition $\A_e\B_e$ of the system. The process is illustrated in Fig.~\ref{fig:TTNS}. \qed

As described for OBC-MPS in Sec.~\ref{sec:oMPS}, any given TTNS with tensors $A_i\in\CC^{d_i\times \prod_{e\in\partial i}m_e}$ can be transformed into an orthonormal form by using the gauge freedom of the tensor network [cf.~Eq.~\eqref{eq:gaugeTrafo}]: To this end, we designate a leaf $i_0$ of the tree as its root and assign directions to all edges according to graph distances to $i_0$ as before. We then visit all vertices $i$ in decreasing order of $\dist(i,i_0)$. Let vertex $i$ have degree $z$, let $e_1$ denote the edge from $i$ to the vertex $i'$ with $\dist(i',i_0)=\dist(i,i_0)-1$. Label the remaining edges from $\partial i$ by $e_2,\dotsc,e_{z}$. Reshape the current tensor $A_i$ into an $m_{e_1}\times(d_i \tilde{m}_{e_2}\dotsb \tilde{m}_{e_z})$ matrix $[A_i]_{n_1,(\s,n_2,\dotsc,n_z)}$, using the index $n_1$ for bond $e_1$ as the row index and $\s\in[1,d_i]$ as well as bond indices $n_2,\dotsc,n_z$ as a multi-index for the columns. Then do a reduced RQ decomposition
\begin{equation}\textstyle
	[A_{i_k}]_{n_1,(\s,n_2,\dotsc,n_z)} \stackrel{\RQ}{=:} \sum_{n'_1=1}^{\tilde{m}_{e_1}} [R]_{n_1,n'_1} [B_i]_{n'_1,(\s,n_2,\dotsc,n_z)}
\end{equation}
and move the matrix $R$ to site $i'$ by contracting it appropriately with tensor $A_{i'}$. Continue with further RQ decompositions. Upon reaching the root $i_0$, the TTNS is in an orthonormal form with all tensors $B_i$ obeying isometry constraints as in Eq.~\eqref{eq:oMPSon} and having (reduced) bond dimensions $\tilde{m}_e\leq m_e$ that agree with the Schmidt ranks. The only remaining freedom in the representation corresponds to unitary gauge transformations \eqref{eq:gaugeTrafo} for each edge. Thus, $\TTNS(\G)$ can be seen as the image of the bounded and closed set
\begin{equation}\label{eq:TTNS-Gr}\textstyle
	\bigcup_{1\leq\tilde{m}_i\leq m_i} \U(1)\bigtimes_i\Gr(\tilde{m}_{e_i},d_i \prod_{e\in\partial i\setminus\{e_i\}} \tilde{m}_e),
\end{equation}
where $e_i$ labels the (unique) edge that starts at vertex $i$. Each term in the union \eqref{eq:TTNS-Gr} is a compact set with real dimension $2\sum_{i=1}^N d_i\prod_{e\in\partial i}\tilde{m}_e-2\sum_{e\in\E}\tilde{m}_e^2 -1$. $\TTNS(\G)$ is the image of the union \eqref{eq:TTNS-Gr} under the continuous (multi-linear) map that contracts the tensors to give the TTNS $|\Psi\ket$. This is an alternative way of seeing that $\TTNS(\G)$ is closed \cite{Rudin1976}.

Concerning the closedness of the Schmidt-rank-bounded set $\R(\G)$, a slightly more general result was presented in Ref.~\cite{Hackbusch2012}, which allows for infinite-dimensional site Hilbert spaces (cf.~Lemma 11.55).

\section{Multiscale entanglement renormalization ansatz}\label{sec:MERA}
\begin{figure*}[t]
\label{fig:MERA}
\includegraphics[width=1\textwidth]{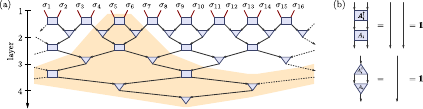}
\caption{(a) A so-called binary MERA for a one-dimensional system with $L$ sites consists of $\log_2 L$ layers. Each layer contains unitary gates that disentangle nearby degrees of freedom before they are decimated by a factor of two. The so-called causal cone of sites 5 and 6 is indicated by the shaded region. (b) All tensors are isometric; cf.\ Eq.~\eqref{eq:MERA-iso}.}
\end{figure*}

MERA states \cite{Vidal-2005-12,Vidal2006} are a hierarchical type of TNS that is inspired by real-space renormalization group schemes \cite{Kadanoff1966-2,Jullien1977-38,Drell1977-16}. In each renormalization step (layer), the system is partitioned into small blocks. Those blocks are to some extent disentangled from neighboring blocks by local unitary transformations before the number of degrees of freedom per block is reduced by application of isometries. A formal characterization of the admissible tensor network structures for a MERA is, for example, given in Ref.~\cite{Barthel2010-105}. Every finite set of sites $\A$ is associated with a \emph{causal cone} that consists of all tensors of the MERA that can affect measurements on subsystem $\A$. The decisive feature of MERA is that the cross-section sizes of such causal cones have a systems-size independent bound. This implies that computation costs for the evaluation of expectation values of strictly local observables scale logarithmically in the system size.
The example of a binary MERA is shown in Fig.~\ref{fig:MERA}.

As for the TTNS, we specify MERA networks $\G=(\E,\{m_e\},\{d_i\})$ by a connected graph with vertices $i=1,\dots,N$ and a set $\E$ of edges, which are associated with site Hilbert spaces of dimension $d_i$ and bond vector spaces of dimension $m_e$. In contrast to TTNS, the MERA network contains loops, and the edges are directed, all pointing in the ``renormalization'' direction. For vertex $i$, $\In(i)$ will denote its incoming edges and $\Out(i)$ its outgoing edges. In a MERA, typically, only the tensors of the first layer carry physical indices $\s_i$, i.e., $d_i=1$ for vertices of all other layers.

A MERA $|\Psi\ket$ is characterized by tensors $A_i\in\CC^{d_i\times m_i\times m_i'}$ associated to the vertices of the graph, where $m'_i:=\prod_{e\in\In(i)}m_{e}$ is the total dimension of the bond vector spaces of incoming edges and $m_i:=\prod_{e\in\Out(i)}m_{e}$ is the one for the outgoing edges. Importantly, all tensors of a MERA are unitaries or isometries. So, when considering $A^{\s_i}_i$ as $m_i\times m'_i$ matrices, mapping from the bond vector spaces of incoming edges to the bond vector spaces of outgoing edges, we have
\begin{equation}\label{eq:MERA-iso}\textstyle
	\sum_{\s=1}^{d_i} A_i^\s A_i^{\s\dag}=\id.
\end{equation}
The MERA $|\Psi\ket$ is obtained by contraction of the tensors $A_i$, i.e., summing over all their bond indices $n_e=1,\dots,m_e$. The set of such normalized MERA states is denoted by $\MERA(\G)$.

According to the isometry constraint \eqref{eq:MERA-iso}, the MERA tensors $A_i$ are elements of the Stiefel manifold $\St(m_i,d_i m'_i)$. Removing a unitary gauge freedom [cf.\ Eq.~\eqref{eq:gaugeTrafo}] in the tensor network at the outgoing edges, they become elements of quotient manifolds
\begin{equation}\textstyle
	\Q_i:=\St(m_i,d_i m'_i)/\left(\bigtimes_{e\in\Out(i)}\U(m_e)\right).
\end{equation}
Thus, $\MERA(\G)$ is the image of the direct product $\U(1)\times\Q_1\times\dotsb\times \Q_N$ under the continuous map that contracts the tensors $A_i$ to give the MERA state $|\Psi\ket$. As the direct product is a compact set, we arrive at the following.
\begin{prop}\label{prop:MERA}
The set $\MERA(\G)$ of normalized MERA states for any network $\G=(\E,\{m_e\},\{d_i\})$ is closed.
\end{prop}

In contrast to OBC-MPS and TTNS, the bond dimensions of a MERA have no one-to-one relation to Schmidt ranks. In fact, they can only be used to derive upper bounds on Schmidt ranks. There is a decisive distinction between MERA in one spatial dimension and higher-dimensional systems. For one-dimensional systems, the Schmidt rank for a block of $\ell$ sites can grow as $m^\ell$ such that MERA can describe critical systems with entanglement entropies $S(\ell)\propto\log\ell$. For higher-dimensional systems, it turns out that MERA form a subclass of PEPS \cite{Barthel2010-105}. The Schmidt rank for a region $\A$ is then bounded by $m^{c|\partial\A|}$, where $|\partial A|$ denotes the surface area of $\A$. Hence, entanglement entropies obey the area law \cite{Eisert2008,Latorre2009,Laflorencie2016-646}. In contrast to PEPS, MERA expectation values for local observables can be evaluated efficiently without approximation, thanks to the isometry constraint \eqref{eq:MERA-iso} and the bounded cross-sections of the causal cones.

\section{Translation-invariant MPS with PBC}\label{sec:tiMPS}
The set $\tiMPS(N,m,d)$ of normalized translation-invariant MPS with PBC contains all states of the form
\begin{equation}\label{eq:tiMPS}
	|\Psi(A)\ket:=\sum_\vs \Tr( A^{\s_1}A^{\s_2}\dotsb A^{\s_N})|\vs\ket\quad\text{with}\quad
	\|\Psi(A)\|=1,
\end{equation}
characterized by a single order-3 tensor $A\in \CC^{d\times m\times m}$. In the following, we will show that $\tiMPS(N,m,d)$ is in general not closed.

To this purpose, consider site Hilbert spaces of dimension $d=2$ and the family of states
\begin{equation}\label{eq:psiW}
	|\psi_W\ket:=\frac{1}{\mc{N}}(|\vphi\ket^{\otimes N}-|1\ket^{\otimes N})
	\quad\text{with}\quad |\vphi\ket:=|1\ket+\veps|2\ket\quad\text{and}\quad
	\mc{N}=\sqrt{(1+\veps^2)^N-1}.
\end{equation}
In the limit $\veps\to 0$, we obtain the so-called \emph{W state} \cite{Duer2000-62}
\begin{equation}\label{eq:Wstate}
	|W\ket:=\lim_{\veps\to 0}|\psi_W\ket=\frac{1}{\sqrt{N}}\left(|2,1,1,1,\dotsc\ket+|1,2,1,1,\dotsc\ket+|1,1,2,1,\dotsc\ket+\dotsc\right),
\end{equation}
which, physically, can be interpreted as a zero-momentum single-magnon state.

Incidentally, this is a well-known example for the fact that sets of tensors with a given tensor rank [a.k.a.\ canonical polyadic (CP) rank] are in general not closed \cite{Bini1979-8,deSilva2008-30}. For a tensor $Q\in\CC^{d_1\times\dotsb\times d_k}$, the tensor rank $r$ is the minimal number of tensor products needed for a CP decomposition $Q=\sum_{\nu=1}^r v_{1,\nu}\otimes\dots\otimes v_{k,\nu}$ with $v_{i,\nu}\in\CC^{d_i}$. In the case of the state \eqref{eq:psiW}, the tensor rank is clearly 2 for any nonzero $\veps$. However, in the limit $\veps\to 0$ we obtain the W state \eqref{eq:Wstate}, which has tensor rank $N$.

For any $\veps>0$, the state \eqref{eq:psiW} can be written as a tiMPS \eqref{eq:tiMPS} with bond dimension $m=2$. Specifically,
\begin{equation}\label{eq:psiWA}
	|\psi_W\ket=|\Psi(A)\ket\in\tiMPS(N,2,2)\quad\text{with}\quad
	A=\frac{1}{\mc{N}^{1/N}}\Bmatrix{|\vphi\ket&0\\ 0&e^{i\pi/N}|1\ket}.
\end{equation}
For brevity, we have written the tensor $A\in\CC^{d\times m\times m}$ as an $m\times m$ matrix of elements in $\H_1\simeq \CC^d$.
\begin{figure*}[t]
\label{fig:tiMPS}
\includegraphics[width=\textwidth]{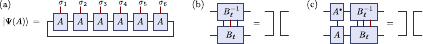}
\caption{(a) In comparison to MPS with OBC \eqref{eq:oMPS}, MPS with PBC \eqref{eq:tiMPS} contain an additional contraction line that connects the tensors of sites 1 and $N$. It allows to encode correlations between the two ends of the chain in a local fashion. In a tiMPS, all tensors are chosen to be identical. (b) After blocking the tensors of $\ell$ consecutive sites, we may arrive at a tensor $B_\ell^{\s_1,\dotsc,\s_\ell}:=A^{\s_1}\dotsb A^{\s_\ell}$ that is injective such that an inverse $B^{-1}_\ell$ exists in the sense of Eq.~\eqref{eq:injectiveBlockTensor}. (c) It follows that $B_{\ell'}$ is injective for all $\ell'\geq\ell$ if $A$ obeys the isometry constraint $\sum_{\s=1}^d A^\s A^{\s\dag}=\id$; cf.~Fig.~\ref{fig:oMPS}b.}
\end{figure*}

Combining results on canonical forms of tiMPS and the injectivity of their tensor factors from Refs.~\cite{Fannes1992-144,PerezGarcia2007-7,Sanz2010-56,Michalek2019-65}, we will see why the limit \eqref{eq:Wstate} is in general not an element of $\tiMPS(N,m,2)$ for sufficiently large $N$ or sufficiently small $m$.
\begin{prop}\label{prop:tiMPS}
The sets $\tiMPS(N,m,d)$ for $m,d\geq 2$ are not closed for prime-number system sizes $N>N_c(m)$, where $2\log_d m< N_c(m)\leq 12 m^3(7+\log_2 m)$.
\end{prop}
The stated bounds on $N_c(m)$ are certainly not tight and can be improved; we will shortly discuss the general case at the end of the section. $\tiMPS(N,1,d)$ is the set of translation-invariant product states, which is clearly closed, and the case $d=1$ is trivial. Let us now go through the proof.

\Emph{Canonical form of tiMPS} \cite{Fannes1992-144,PerezGarcia2007-7} -- Translation invariant states of a system with $N$ sites and PBC can always be written as tiMPS $|\Psi(A)\ket$, and the tensor $A$ can always be brought into the canonical form
\begin{equation}\label{eq:tiMPS-c}
	A^\s=\Bmatrix{\alpha_1 A^\s_1 & &\\ & \ddots & \\ & & \alpha_b A^\s_b}\quad \text{for}\quad
	\s=1,\dotsc,d.
\end{equation}
with $b$ blocks on the diagonal. $A^\s_j$ has dimension $m_j\times m_j$, $0<\alpha_j\leq 1$, and $\sum_\s A^\s_j A^{\s\dag}_j=\id_{m_j}$ such that $\{A^{\s\dag}_j\,|\,\s=1,\dots,d\}$ can be interpreted as Kraus operators that define the quantum channel $\E_j(X):=\sum_\s A^{\s\dag}_j X A^\s_j$ \cite{Nielsen2000}. For the canonical form, these channels have unique diagonal fixed-point matrices $\Lambda_j\succ 0$ with $\E_j(\Lambda_j)=\Lambda_j$. If one has a tiMPS $|\Psi(\tilde{A})\ket$ with bond dimension $m$, the corresponding canonical tensor \eqref{eq:tiMPS-c} with $|\Psi(A)\ket=|\Psi(\tilde{A})\ket$ has bond dimension $\sum_{j=1}^b m_j\leq m$.

\Emph{Injective MPS and quantum Wielandt's inequality} \cite{Sanz2010-56,Michalek2019-65} -- An MPS tensor $A\in\CC^{d\times m\times m}$ that obeys the isometry constraint $\sum_\s A^\s A^{\s\dag}=\id$ may give rise to an \emph{injective} tensor
\begin{equation}\label{eq:l-siteTensor}
	B_\ell^{\s_1,\dotsc,\s_\ell}:=A^{\s_1}A^{\s_2}\dotsb A^{\s_\ell}
\end{equation}
that is obtained by contracting the tensors of $\ell$ consecutive sites; i.e, the $\ell$-site tensor $B_\ell$,
seen as a map from the joint bond vector space $\CC^{m\times m}$ to the $\ell$-site Hilbert space $\H_1^{\otimes\ell}$, may be injective \cite{Fannes1992-144,PerezGarcia2007-7,Molnar2018-20}. Injectivity of $B_\ell$ is equivalent to requiring that $\{B_\ell^\vs\,|\,\vs\in [1,d]^{\times \ell}\}$ is a basis for the $m^2$-dimensional space of $m\times m$ matrices. Let $B_\ell$ be injective with an inverse $B^{-1}_\ell$ such that
\begin{equation}\label{eq:injectiveBlockTensor}\textstyle
	\sum_\vs [(B^{-1}_\ell)^\vs]_{n_1,n_2} [B^\vs_\ell]_{n_1',n_2'}=\delta_{n_1,n_1'}\delta_{n_2,n_2'}.
\end{equation}
Then $(A^{\s'})^\ast (B^{-1}_\ell)^\vs $ is an inverse of $B^{\s',\vs}_{\ell+1}$, i.e., $B_{\ell'}$ is injective for all $\ell'\geq \ell$; cf.\ Fig.~\ref{fig:tiMPS}. Furthermore, for a canonical tiMPS tensor \eqref{eq:tiMPS-c} with blocks $j=1,\dotsc,b$, the $\ell'$-site subspaces
\begin{equation}\label{eq:l-siteTensorBlock} \textstyle
	\H_j:=\Span\big\{\sum_{\vs} [B_{j,\ell'}^{\vs}]_{n,n'}|\vs\ket\,|\,n,n'\in [1,m]\big\}\quad\text{with}\quad
	B_{j,\ell}^{\s_1,\dotsc,\s_\ell}:=A_j^{\s_1}\dotsb A_j^{\s_\ell}
\end{equation}
are mutually orthogonal for $\ell'\geq 3(b-1)(\ell+1)$ if the tensors $B_{j,\ell}$ are injective for all $j$ \cite{PerezGarcia2007-7}.

Now, a quantum version of Wielandt's inequality was proven in Ref.~\cite{Sanz2010-56}. It states that, if there exists an $\ell$ for which $B_\ell$ is injective, then it will be injective for all $\ell \geq m^4$. An improved upper bound $N_\inj(m)$ on the \emph{injectivity length} was given in Ref.~\cite{Michalek2019-65} with
\begin{equation}\label{eq:WielandtIneq}
	\ell\geq N_\inj(m):= 2 m^2(6+\log_2 m).
\end{equation}

As discussed above, the canonical form of the tiMPS \eqref{eq:tiMPS-c} gives rise to a quantum channel $\E_j$ for each block $j\in[1,b]$. Each channel $\E_j$ has a unique fixed point $\Lambda_j$, i.e., its eigenvalue 1 is non-degenerate. For this case, Fannes \emph{et al.} \cite{Fannes1992-144} showed that all eigenvalues of $\E_j$ of magnitude 1 are roots $e^{2\pi i n/p}$ of the identity for some $p\in\NN$ and $n=1,\dotsc,p$. There also exists a set of $p$ orthogonal projectors $P_n$ with $\sum_{n=1}^p P_n=\id_{m_j}$ and $P_n A_j^\s =A_j^\s P_{n+1}$, where $P_{p+1}\equiv P_1$. It follows that the contribution of block $j$ in the canonical form vanishes unless $p$ is a divisor of the system size $N$ \cite{PerezGarcia2007-7}. Thus, if we restrict our considerations to prime-number $N$, 1 is the only magnitude-1 eigenvalue of the $\E_j$ and the corresponding fixed-point matrices $\Lambda_j$ are positive. Quantum channels of this type are \emph{primitive}. Repeated application of a primitive channel $\E_j$ leads exponentially fast to the (full-rank) fixed point, and the Kraus operator products $B_{j,\ell}$ span the space of $m_j\times m_j$ matrices for some finite $\ell$, i.e., the $B_{j,\ell}$ are injective and the precondition of the quantum Wielandt's inequality is fulfilled \cite{Sanz2010-56}.

\Emph{Bond-dimension bound for the W state} -- Assume that there exists a tiMPS representation \eqref{eq:tiMPS} of $|W\ket$ in Eq.~\eqref{eq:Wstate} with prime $N$, bond dimension $m$, and MPS tensor $A$ in the canonical form \eqref{eq:tiMPS-c}. According to the Wieldandt inequality \eqref{eq:WielandtIneq}, the $\ell$-site tensors $B_{j,\ell}$ are injective if $\ell\geq N_\inj(m_j)$. Choose $\ell_0=\max_j N_\inj(m_j)$ such that all $B_{j,\ell\geq\ell_0}$ are injective. By contradiction as in Ref.~\cite{PerezGarcia2007-7}, we will see in the following that no such tiMPS representation of the W state can exist for $N\geq 6(m-1)(\ell_0+1)$.

For the given setup, let us choose $N\geq 6(b-1)(\ell_0+1)$ such that we can split the system into two groups $\A$ and $\B$ of $i,N-i\geq 3(b-1)(\ell_0+1)$ consecutive sites, respectively. We can write the W state as a sum of tensor products $|\Psi(A)\ket=\sum_{j=1}^b\sum_{n,n'=1}^{m_j}|j,n,n'\ket_\A\otimes|j,n,n'\ket_\B$, where, similar to Eq.~\eqref{eq:oMPSsplit},
\begin{subequations}\label{eq:tiMPSsplit}
\begin{alignat}{7}
	&|q\ket_\A&&:=|j,n,n'\ket_\A&&=\textstyle\sum_{\s_1,\dotsc,\s_i}     &&[A_j^{\s_{1}}    \dotsb A_j^{\s_i}]_{n,n'}\,&&|\s_{1},  \dotsc,\s_i\ket\quad\text{and}\\
	&|q\ket_\B&&:=|j,n,n'\ket_\B&&=\textstyle\sum_{\s_{i+1},\dotsc,\s_N} &&[A_j^{\s_{i+1}}\dotsb A_j^{\s_N}  ]_{n',n}\,&&|\s_{i+1},\dotsc,\s_N\ket.
\end{alignat}
\end{subequations}
According to the properties of injective tiMPS discussed above [cf.\ Eq.~\eqref{eq:l-siteTensorBlock}],
\begin{equation}
	\Span\{|j,n,n'\ket_\A\,|\,j\in[1,b],\,n,n'\in[1,m_j]\}\quad\text{and}\quad
	\Span\{|j,n,n'\ket_\B\,|\,j\in[1,b],\,n,n'\in[1,m_j]\}
\end{equation}
have both (full) dimension $\sum_{j=1}^b m_j^2$, which is hence the rank of the reduced density matrix
\begin{equation}\label{eq:tiMPS-W-rhoA}\textstyle
	\Tr_\B|\Psi(A)\ket\bra\Psi(A)|=\sum_{q,q'}|q\ket_{\A\,\B}\bra q'|q\ket_{\B\,\A}\bra q'|.
\end{equation}
We know that the reduced density matrix of the W state \eqref{eq:Wstate} for subsystem $\A$ has rank 2, with the support spanned by $|1\ket^{\otimes i}$ and the W state on $i$ sites. This would imply that the canonical tiMPS tensor $A$ has $b=2$ blocks with $m_1=m_2=1$. This is a contradiction since the W state can clearly not be written as a sum of only two product states.

\Emph{System-size bounds for non-closedness} --
The W state is the $\veps\to 0$ limit of the family of tiMPS \eqref{eq:psiWA} with bond dimension $2$, but we have found that $N< 6(m-1)(\ell_0+1)$ for any tiMPS representation with bond dimension $m$. Using that $\ell_0\leq N_\inj(m)$ with Eq.~\eqref{eq:WielandtIneq}, we arrive at the result that $\tiMPS(N,2,2)$ is not closed for prime $N> 12 m^3(7+\log_2 m)$ with $m=2$. The same criteria guarantee non-closedness of $\tiMPS(N,m,d)$ with $m,d>2$, since the states $|\psi_W\ket$ in Eqs.~\eqref{eq:psiW} and \eqref{eq:psiWA} can of course also be written as a tiMPS with bond dimension $m>2$, e.g., by padding the tensor $A$ in Eq.~\eqref{eq:psiWA} with zeros. Moreover, we can allow $d>2$, since it is not necessary for $|\psi_W\ket$ to exhaust the single-site Hilbert spaces.

The lower bound $2\log_d m$ on $N_c(m)$ stated in Prop.~\ref{prop:tiMPS} follows from the fact that, as discussed in Sec.~\ref{sec:oMPS}(c), every state with Schmidt ranks $\{m_i\}$ for bipartitions into subsystem $\A=[1,i]$ and its complement $\B=[i+1,N]$ has an OBC-MPS representation \eqref{eq:oMPS} with tensors $A_i\in\CC^{d\times m_{i-1}\times m_i}$. The maximum possible Schmidt rank for a state in $\H=\H_1^{\otimes N}$ is $\tilde{m}=d^{\lfloor N/2\rfloor}$. From the OBC-MPS representation with tensors $A_i$, we can get a tiMPS representation \eqref{eq:tiMPS}, by choosing $A^\s=N^{-1/N}\sum_{i=1}^N A_i^\s \otimes E_{i,i+1}$. Here, $E_{i,i+1}$ are $N\times N$ matrices with a single nonzero entry $[E_{i,i+1}]_{j,j'}=\delta_{i,j}\delta_{j',i+1\,\mathrm{mod}\,N}$. Thus, every state has a tiMPS representation with bond dimension $m\leq N \tilde{m}=N d^{\lfloor N/2\rfloor}$, i.e., non-closedness requires at least $N>2\log_d m$.
\qed

\Emph{Comments} -- The large injectivity length \eqref{eq:WielandtIneq} and the requirement for prime $N$ are a worst-case scenario. In generic cases, canonical tiMPS representations have primitive block channels $\E_j$, and they are injective for blocks $\ell \geq 2 \log_d m$. Such a generic tiMPS representation of the W state \eqref{eq:Wstate} cannot exist for $N>12 m\log_d m$. Furthermore, note that the requirement of a translation-invariant MPS format \eqref{eq:tiMPS} is essential here: The W state can easily be written as a heterogeneous OBC-MPS \eqref{eq:oMPS} with bond dimension 2, where
\begin{equation}\label{eq:psiWA-obc}
	A_1=\frac{1}{\mc{N}}\Bmatrix{|2\ket&|1\ket},\quad
	A_i=\Bmatrix{|1\ket& 0\\|2\ket&|1\ket} \ \text{for} \ 1<i<N, \ \text{and}\quad 
	A_N=\Bmatrix{|1\ket\\|2\ket}.
\end{equation}

\section{Heterogeneous MPS with PBC}\label{sec:pMPS}
The set $\pMPS(N,m,d)$ of (heterogeneous) MPS with PBC contains all states of the form
\begin{equation}\label{eq:pMPS}
	|\Psi\ket=\sum_\vs \Tr \left( A_1^{\s_1}A_2^{\s_2}\dotsb A_N^{\s_N}\right)|\vs\ket,
\end{equation}
characterized by $N$ order-3 tensors $A_i\in \CC^{d\times m\times m}$. While we consider a fixed single-site dimension $d$ in Sec.~\ref{sec:pMPSsmall}, we generalize to site dependent dimensions $d_i$ in Sec.~\ref{sec:pMPSlarge}.
For clarity of notation, we consider the case where all bond dimensions are equal to $m$, although it is not essential. Note that, so far, we did not impose a normalization constraint. The set of pMPS \eqref{eq:pMPS} with norm $\|\Psi\|=1$ will be denoted by $\pMPSn(N,m,d)$.

\subsection{Small bond dimensions \texorpdfstring{$d\geq m^2$}{d>=m*m}}\label{sec:pMPSsmall}
Landsberg \emph{et al.}~\cite{Landsberg2012-12} discussed the case of small bond dimensions with $\dim\H_1=d\geq m^2$, finding that sets of corresponding heterogeneous MPS with PBC are not closed based on comparing dimensions of stabilizers. In the following, we will rephrase the arguments of Ref.~\cite{Landsberg2012-12} and extend them.

Let us choose $d= m^2$ and impose a tensor-product structure for the single-site Hilbert space $\H_1=\V\otimes\V=\Span\{|n,n'\ket\,|\,n,n'=1,\dotsc,m\}$, where vector spaces $\V$ agree with the bond vector space and are spanned by an orthonormal basis $\{|n\ket\,|\,n=1,\dotsc,m\}$. In this setting, we can write all MPS \eqref{eq:pMPS} as images of the state
\begin{equation}\label{eq:pMPSmuDef}
	|\mu\ket:=\sum_{n_0,\dotsc,n_{N-1}=1}^m|n_0, n_1\ket\otimes|n_1, n_2\ket\otimes |n_2, n_3\ket\otimes\dotsb\otimes |n_{N-1}, n_0\ket
	= \bigotimes_{e}\left(\sum_{n=1}^{m} |n, n\ket_e\right).
\end{equation}
under the application of operators $\hat{A}_i\in\End(\H_1)$ on the site Hilbert spaces such that
\begin{equation}\label{eq:pMPSmu}
	|\Psi\ket=\hat{A}_1\otimes\hat{A}_2\otimes\hat{A}_3\otimes\dotsb\otimes\hat{A}_N|\mu\ket.
\end{equation}
The tensor product on the right-hand side of Eq.~\eqref{eq:pMPSmuDef} runs over all edges $e=(i,i+1)$ of the one-dimensional lattice.
The generated state $|\Psi\ket$ is equal to the MPS \eqref{eq:pMPS} if the matrix elements of the operators $\hat{A}_i$ agree with the elements of the MPS tensors in the sense that $\bra \sigma_i|\hat{A}_i|n_{i-1}n_i\ket=[A_i^{\sigma_i}]_{n_{i-1},n_i}$. Note that this is a standard way of constructing MPS or PEPS \cite{Affleck1988-115,Verstraete2004-7}: One associates copies of each bond vector space to both of the corresponding lattice sites, prepares a tensor product $|\mu\ket$ with each bond in a maximally entangled state $\sum_{n=1}^m|n, n\ket$ (a.k.a.\ entangled-pair state), and applies to it then a tensor product of linear operators $\hat{A}_i$ that map into the actual site Hilbert spaces. The special case considered here is that the single-site Hilbert space is isomorphic to the tensor product of the associated bond vector spaces ($d=m^2$).
\begin{figure*}[t]
\label{fig:pMPS}
\includegraphics[width=0.99\textwidth]{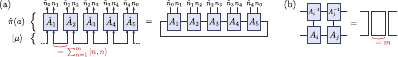}
\caption{(a) The heterogeneous MPS with PBC \eqref{eq:pMPS} considered in Sec.~\ref{sec:pMPSsmall} have small bond dimensions $m$ (large $d$), obeying $d=m^2$. They can be constructed from a tensor product $|\mu\ket$ of maximally entangled states $\sum_n|n,n\ket$ for each bond and subsequent application of endomorphisms $\hA_i$ on every site; cf.\ Eq.~\eqref{eq:pMPSmu}. (b) If consecutive operators $\hA_i$ are injective maps, also their contraction is injective (invertible); cf.\ Eq.~\eqref{eq:ipMPS-inverse} and Fig.~\ref{fig:tiMPS}b.}
\end{figure*}

\Emph{Injective pMPS as a group orbit} --
Consider the sets $\E$ and $\G$ of $N$-fold tensor products of generic single-site operators $\End(\H_1)$ and invertible operators $\GL(\H_1)$,
\begin{subequations}\label{eq:pMPS_E_G}
\begin{gather}
	\E:=\{\hat{A}_1\otimes\dotsb\otimes\hat{A}_N\,|\,\hat{A}_i\in\End(\H_1)\}\quad\text{and}\\
	\G:=\{\hat{A}_1\otimes\dotsb\otimes\hat{A}_N\,|\,\hat{A}_i\in\GL(\H_1)\}\subset\E,
\end{gather}
\end{subequations}
respectively.
According to Eq.~\eqref{eq:pMPSmu}, we then have
\begin{equation}\label{pMPS-setDef}
	\pMPS(N,m,m^2)=\E|\mu\ket
	\equiv\{\hat{a}|\mu\ket\,|\,\hat{a}\in\E\}.
\end{equation}
The subset of injective MPS is
\begin{equation}\label{ipMPS-setDef}
	\ipMPS(N,m,m^2):=\G|\mu\ket
	\equiv\{\hat{a}|\mu\ket\,|\,\hat{a}\in\G\}.
\end{equation}
These are injective in the sense that the considered MPS operators $\hA_i\in\GL(\H_1)$ correspond to injective maps from the bond vector spaces $\V\otimes\V$ to the single-site Hilbert space $\H_1$. This is precisely the notion of injectivity discussed in Sec.~\ref{sec:tiMPS} with injectivity length 1.

In contrast to $\E$, $\G$ is a group with the operator multiplication on $\H$ as the group operation. It is a Lie group that is generated by the identity and all traceless operators on site $i$, tensored with the identity on the remaining sites, where $i=1,\dots, N$. Hence,
\begin{equation}\label{eq:pMPS-dimG}
	\dim_\CC \G=N\cdot(m^4-1)+1.
\end{equation}
$\ipMPS$ has very nice properties. In particular, it is the orbit of the state $|\mu\ket$ in Eq.~\eqref{eq:pMPSmuDef} under the action of the group $\G$. Orbits are homogeneous spaces, i.e., for all $|\Psi\ket,|\Phi\ket\in\ipMPS$ there exists a group element $\hat{a}\in\G$ such that $|\Phi\ket=\hat{a}|\Psi\ket$.
Moreover, the constraint $\det\hat{A}_i\neq 0$ for invertible $\hat{A}_i$ does not reduce the dimension of the operator set as a quasi-projective variety such that
\begin{equation}\label{eq:pMPS-dim0}
	\dim \G=\dim\E\quad\text{and, hence,}\quad
	\dim \ipMPS(N,m,m^2)=\dim \pMPS(N,m,m^2).
\end{equation}
Finally, the constraints $\det\hat{A}_i\neq 0$ are not stable with respect to taking the set closure. Thus,
\begin{equation}
	\overline{\ipMPS(N,m,m^2)}=\overline{\pMPS(N,m,m^2)}.
\end{equation}

\Emph{Dimension of pMPS} -- According to Eq.~\eqref{eq:pMPS-dim0} we can study the dimension of $\pMPS$ through that of $\ipMPS$. From group theory, we know that 
\begin{equation}\label{eq:pMPS-dim1}
	\dim \ipMPS\equiv\dim \G|\mu\ket=\dim\G-\dim\G_\mu,\quad \text{where}\quad
	\G_\mu:=\{\hat{a}\in\G\,|\,\hat{a}|\mu\ket=\mu\ket\}
\end{equation}
is the \emph{stabilizer} of the state $|\mu\ket$ under the action of $\G$. The entangled-pair states $\sum_{n=1}^m|n, n\ket$ are invariant under transformations of the form
\begin{equation}\label{eq:pMPS-Zinv}
	\hZ^{-1}\otimes\hZ^T\sum_{n=1}^m|n, n\ket
	=\sum_{n,n',n''=1}^m|n',n''\ket\bra n'|\hZ^{-1}|n\ket\bra n''|\hZ^T|n\ket
	=\sum_{n=1}^m|n,n\ket
\end{equation}
where the transposition is defined such that $\bra n''|\hZ^T|n\ket\equiv \bra n|\hZ|n''\ket$. Hence,
\begin{equation}\label{eq:pMPS-Gmu}
	\G_\mu=\{\hat{z}_1\otimes\dotsb\otimes\hat{z}_N\,|\,\hat{z}_i=\hZ^T_{i-1}\otimes\hZ^{-1}_i,\, \hZ_i\in\SL(\V)\}.
\end{equation}
Here, we employ the special linear group $\SL(\V)$ to fix a scale-freedom of the $\hZ_i$ through $\det\hZ_i=1$. So,
\begin{equation}\label{eq:pMPS-Gmu-dim}
	\dim_\CC \G_\mu=N m^2-N,
\end{equation}
where the reduction by $N$ is due to the constraints on the determinants for $i=1,\dotsc,N$. With Eqs.~\eqref{eq:pMPS-dimG} and \eqref{eq:pMPS-dim1} we obtain
\begin{equation}
	\dim_\CC\pMPS(N,m,m^2)=N m^2(m^2-1)+1.
\end{equation}
Note that the invariance of $|\mu\ket$ under the action of the stabilizer \eqref{eq:pMPS-Gmu} is in fact nothing but the usual MPS gauge freedom \eqref{eq:gaugeTrafo} as
\begin{equation}\textstyle
	\bra\s|\hA_i(\hZ^T_{i-1}\otimes\hZ^{-1}_i)|n,n'\ket
	=\sum_{\bar{n},\bar{n}'=1}^m[A^\sigma_i]_{\bar{n},\bar{n}'}\bra \bar{n}|\hZ^T_{i-1}|n\ket\bra\bar{n}'|\hZ^{-1}_i|n'\ket
	=[Z_{i-1}A^\sigma_iZ^{-1}_{i}]_{n,n'}.
\end{equation}

\begin{figure*}[t]
\label{fig:2DSstabilizer}
\includegraphics[width=0.83\textwidth]{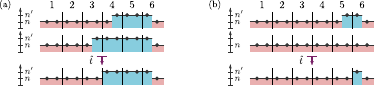}
\caption{The two-domain state \eqref{eq:2DS} is an equal-weight superposition of tensor product states that have two domains with sites in basis states $|n\ket$ and $|n'\ket$, respectively, where $n,n'\in [1,m]$. The domain walls are located on sites $i\in[1,N-1]$ and $N$, respectively. Two components are depicted for $N=6$ in the upper parts of both panels (a) and (b). The off-diagonal generators for the stabilizer of the two-domain state $|\tau\ket$ are spanned by operators $\hT$ in Eq.~\eqref{eq:2DSstabilizerOD} that act on neighboring sites to map two such components into the same state (bottom configurations) but with opposite sign such that $\hT|\tau\ket=0$.}
\end{figure*}

\Emph{Two-domain state} -- Following Ref.~\cite{Landsberg2012-12}, consider the family of injective pMPS for $\veps>0$
\begin{subequations}\label{eq:psiTau}
\begin{align}
	&|\psi_\tau\ket:=\hat{A}\otimes\hat{A}\otimes\dotsb\otimes\hat{A}\otimes\hat{B}|\mu\ket\in\ipMPS\quad\text{with}\quad
	\hA=\hP_d+\veps\hP_o,\quad
	\hB=\frac{1}{\veps}(\hP_o+\veps\hP_d),\\
	&\hP_d:=\sum_{n=1}^m|n,n\ket\bra n,n|,\quad
	\hP_o:=\sum_{n=1}^m\sum_{n'\neq n=1}^m|n,n'\ket\bra n,n'|=\id-\hP_d,
\end{align}
\end{subequations}
i.e., $\hat{A}=\veps \id+(1-\veps)\hP_d$. Noting that $\hP_d^{\otimes (N-1)}\otimes\hP_o|\mu\ket=0$, the limit $\veps\to 0$ is well-defined, and we obtain the state
\begin{align}\nonumber
	|\tau\ket
	&:= \lim_{\veps\to 0}|\psi_\tau\ket
	=\hP_d^{\otimes N}|\mu\ket+\sum_{i=1}^{N-1}\hP_d^{\otimes (i-1)}\otimes\hP_o\otimes\hP_d^{\otimes (N-i-1)}\otimes \hP_o|\mu\ket\\
	\label{eq:2DS}
	&=\sum_n|n,n\ket^{\otimes N} + \!\!\sum_{i,n,n'\neq n}\!\!|n,n\ket_1\otimes|n,n\ket_2\otimes\dotsb\otimes|n,n'\ket_i\otimes |n',n'\ket_{i+1}\otimes\dotsb\otimes|n',n\ket_{N},
\end{align}
where the subscripts in the second line indicate lattice sites. This is a sum of product states with a domain wall at site $i\in[1,N-1]$ and another at site $N$. Hence, we call $|\tau\ket$ the \emph{two-domain state}; see Fig.~\ref{fig:2DSstabilizer}. With $\A_j:=\{j\}$ and $\B_j:=[1,N]\setminus\{j\}$, we have a bipartition of the system into site $j$ and the rest. One finds that the reduced density matrices $\dm_j:=\Tr_{\B_j}|\tau\ket\bra\tau|$ all have full rank, $\Rank\dm_j=m^2$ $\forall_j$: For $j<N$, we can write
\begin{align*}
	&\hspace{-4ex}\textstyle|\tau\ket=\sum_n\Big( |n,n\ket_j\otimes |n,n\ket_{\B_j}+\sum_{n'\neq n}|n,n'\ket_j\otimes|n,n'\ket_{\B_j}\Big)\quad\text{with}\\
	&\hspace{-4ex}\textstyle|n,n\ket_{\B_j}=|n,n\ket^{\otimes(N-1)}+\sum_{n'\neq n}\big(\sum_{i>j}|n,n\ket_{j+1}\otimes\dotsb\otimes |n,n'\ket_i\otimes\dotsb\otimes|n',n\ket_N\otimes\dotsb\otimes|n,n\ket_{j-1}\\
	&\hspace{-4ex}\textstyle\hspace{28.8ex} +\sum_{i<j}|n,n\ket_{j+1}\otimes\dotsb\otimes |n,n'\ket_N\otimes\dotsb\otimes|n',n\ket_i\otimes\dotsb\otimes|n,n\ket_{j-1}\big),\\
	&\hspace{-4ex}\textstyle|n,n'\ket_{\B_j}=|n',n'\ket_{j+1}\otimes\dotsb\otimes |n',n\ket_N\otimes\dotsb\dotsb\otimes|n,n\ket_{j-1}.
\end{align*}
All states for site $j$ and block $\B_j$ in this decomposition are mutually orthogonal such that this is, up to normalization, a Schmidt decomposition of $|\tau\ket$ with $m^2$ terms, which implies $\Rank\dm_j=m^2$. The same works for $j=N$.

\Emph{Location of the two-domain state} -- 
The injectivity of $|\Psi\ket\in\ipMPS$ [cf.~Eq.~\eqref{ipMPS-setDef}] implies that all single-site density matrices $\dm_j=\Tr_{\B_j}|\Psi\ket\bra\Psi|$ have full rank, $\Rank\dm_j=m^2$: As discussed in Sec.~\ref{sec:tiMPS}, injectivity of the operator $\hA_j$ implies that $\{A^\s_j\,|\,\s\in[1,d]\}$ spans the space of $m\times m$ matrices. Furthermore, the injectivity of $\hA_i$ and $\hA_{i+1}$ implies that $\hB:=(\hA_i\otimes \hA_{i+1})\sum_{n,n',n''}|n,n',n',n''\ket\bra n,n''|$ is injective. In particular, with the inverses $\hA^{-1}_i$ and $\hA^{-1}_{i+1}$ of the single-site tensors, 
\begin{equation}\label{eq:ipMPS-inverse}
 	\hB^{-1}=\frac{1}{m}\sum_{n,n',n''}| n,n''\ket\bra n,n',n',n''|(\hA^{-1}_i\otimes \hA^{-1}_{i+1})
\end{equation}
is the inverse of $\hB$; cf.\ Fig.~\ref{fig:pMPS}. Consequently, also $\{A^{\s_{j+1}}_{j+1}\dotsb A^{\s_{N}}_{N}A^{\s_{1}}_{1}\dotsb A^{\s_{j-1}}_{j-1}\,|\,\s_i\in[1,d]\}$ spans the space of $m\times m$ matrices. In the decomposition
\begin{equation}
	|\Psi\ket=\sum_{n,n'=1}^m|n,n'\ket_j\otimes |n,n'\ket_{\B_j}
\end{equation}
of the ipMPS $|\Psi\ket$, in analogy to Eqs.~\eqref{eq:tiMPSsplit}-\eqref{eq:tiMPS-W-rhoA}, we have two sets of $m^2$ linearly independent states and, hence, $\Rank \dm_j=m^2$.

As all single-site density matrices $\dm_j$ of the two-domain state \eqref{eq:2DS} have full rank, there are two possibilities. Either $|\tau\ket$ is in $\ipMPS(N,m,m^2)$, or it is an element of the boundary $\partial\pMPS$ of $\pMPS$ that is not contained in $\pMPS$. It cannot be in $\pMPS\setminus \ipMPS$ because all those states have at least one operator $\hA_j$ that is not injective and hence a $\dm_j$ without full rank. Comparing the dimension of the stabilizer $\G_\tau$ with that of $\G_\mu$, we will see in the next subsection that the latter scenario is the case and that, hence, $\pMPS(N,m,m^2)$ is not a closed set.

\Emph{Stabilizer of the two-domain state} -- 
If the two-domain state \eqref{eq:2DS} were in $\ipMPS$, the dimension of its stabilizer $\G_\tau=\{\hat{a}\in\G\,|\,\hat{a}|\tau\ket=|\tau\ket\}$ would agree with that of $\G_{\mu}$ because, as previously discussed, $\ipMPS$ is a homogeneous space with respect to the group $\G$ [Eq.~\eqref{eq:pMPS_E_G}]. Generators for elements of $\G_\tau$ take the form $\hT=c\id+\sum_{j=1}^N\hT_j$ with operators $\hT_j$ acting non-trivially on site $j$ and as the identity on all other sites. They obey $\hT|\tau\ket=0$. Off-diagonal $\hT$ necessarily act simultaneously on neighboring sites: For $N\geq 3$, they either map the equal-weight sum of two terms with domain walls on sites $i$ and $i-1$, respectively, to zero, or they map the sum of a term with a domain wall on site $1$ or $N-1$ and a term without domain walls to zero; see Fig.~\ref{fig:2DSstabilizer}. Thus, a basis for the off-diagonal generators is given by
\begin{equation}\label{eq:2DSstabilizerOD}
\begin{split}
	&|n',n'\ket\bra n,n'|_i - |n,n\ket\bra n,n'|_{i-1}\quad\text{for} \ \ 1<i<N,\quad\text{and}\\
	&|n',n'\ket\bra n,n'|_1 - |n',n\ket\bra n',n'|_{N},\quad
	|n',n\ket\bra n,n|_N - |n,n\ket\bra n,n'|_{N-1}.
\end{split}
\end{equation}
The subscripts indicate sites and tensoring with the identity on the remaining sites is implied. See Fig.~\ref{fig:2DSstabilizer}. With $N$ such operators for each index pair $n\neq n'\in[1,m]$, this gives $Nm(m-1)$ linearly independent off-diagonal generators. Generators $\hT$ that are diagonal in the basis $\{|n,n'\ket\}$ need to solve the equations
\begin{equation}
\begin{split}
	&\textstyle\sum_{j=1}^{i-1}\bra n,n|\hT_j|n,n\ket+\bra n,n'|\hT_i|n,n'\ket+\sum_{j= i+1}^{N-1}\bra n',n'|\hT_j|n',n'\ket+\bra n',n|\hT_N|n',n\ket=-c,\\
	&\textstyle\sum_{j=1}^N\bra n,n|\hT_j|n,n\ket = -c
\end{split}
\end{equation}
for all $i\in[1,N-1]$ and $n\neq n'\in[1,m]$. With $N\cdot(m^2-1)+1$ diagonal matrix elements and, $(N-1)m(m-1)+m$ independent equations, this gives $N\cdot(m-1)+m(m-2)+1$ linearly independent diagonal generators. In summary, for $N\geq 3$,
\begin{equation}\label{eq:2DSstabilizerDim}
	\dim_\CC \G_\tau = Nm(m-1) + N\cdot(m-1)+m(m-2)+1 =  N(m^2-1)+m(m-2)+1.
\end{equation}
This is consistent with the value $4m^2-2m$ given in Ref.~\cite{Landsberg2012-12} for $N=3$ because, in that paper, the group $\G$ was defined in terms of a Cartesian product instead of the tensor product in Eq.~\eqref{eq:pMPS_E_G}. Also note that Eq.~\eqref{eq:2DSstabilizerDim} correctly captures the corner case $m=1$, for which $\G_\tau=\{\id\}$, i.e., $\dim\G_\tau=0$. 
For any $N\geq 3$ and $m>1$, the obtained $\dim_\CC \G_\tau$ is larger than $\dim_\CC \G_{\mu}$ [Eq.~\eqref{eq:pMPS-Gmu-dim}]. Hence, the two-domain state \eqref{eq:2DS} is not in $\pMPS(N,m,m^2)$ and we have established that $\pMPS(N,m,m^2)$ is not a closed set.

\Emph{Normalization} -- 
So far, we did not impose a normalization constraint on the MPS. However, the considered states $|\psi_\tau\ket$ and their $\veps\to 0$ limit $|\tau\ket$ are normalizable. Enforcing norm one, reduces the real dimension of the MPS quasi-projective variety by one ($\dim_\RR \pMPSn=\dim_\RR \pMPS -1$), but it remains a non-closed set: All $|\psi_\tau\ket/\|\psi_\tau\|$ are in $\pMPSn$, and normalization of $|\tau\ket$ does not change the fact that it cannot be represented as a pMPS with bond dimension $m$.
\begin{prop}\label{prop:pMPSsmall}
The sets $\pMPSn(N,m,d\geq m^2)$ of normalized pMPS with bond dimension $m>1$ on a ring of $N$ sites are not closed for any $N\geq 3$.
\end{prop}
Here, we generalized from $d=m^2$ to $d\geq m^2$ because, for the above arguments to hold, it is not necessary for the states $|\psi_\tau\ket$ and $|\tau\ket$ to exhaust the single-site Hilbert spaces.

\subsection{Extension to larger bond dimensions}\label{sec:pMPSlarge}
In actual simulations of quantum many-body systems, bond dimensions $m$ are typically much larger than single-site Hilbert space dimensions $d$, say, $d\sim 2\dots 10$ and $m\sim 10\dots 15000$. Often, one is interested in simulating large system sizes $N$.
The result on non-closedness can be generalized to this more relevant regime by imposing a tensor-product structure on the local Hilbert spaces and decomposing the tensors of the injective pMPS \eqref{eq:psiTau} to obtain an MPS for a correspondingly fine-grained lattice.
\begin{prop}\label{prop:pMPSbig}
For a system with size $N$ being a multiple of $\ell\in\NN$ and some single-site Hilbert-space dimension $\tilde{d}\geq d > 1$, the sets $\pMPSn(N,m=d^{\ell/2},\tilde{d})$ are not closed for any $N\geq 3\ell$.
More generally, if the system consists of at least three clusters with lengths $\ell_c$ such that $N=\sum_c\ell_c$, and with some single-site Hilbert space dimensions $\tilde{d}_{c,1}\geq d_{c,2},\dotsc,\tilde{d}_{c,\ell_c}\geq d_{c,\ell_c}$ in cluster $c$ such that $1<m^2=d_1 d_2 \dotsb d_{\ell_c}$, then the corresponding set $\pMPSn(N,m,\{\tilde{d}_{c,i}\})$ is not closed.
\end{prop}

\Emph{Proof:}
\textbf{(a)} In a first step, we decompose the tensors
\begin{equation}\label{eq:psiTauA}\textstyle
	\hA=\alpha \hP_d+\beta\hP_o
	=\beta \id+(\alpha-\beta)\sum_{n=1}^m|n,n\ket\bra n,n| \quad\text{with}\quad 
	\alpha,\beta\in\RR
\end{equation}
for the MPS representation \eqref{eq:psiTau} of $|\psi_\tau\ket$, where $\alpha=1$ and $\beta=\veps$ for the tensors of the first $N-1$ sites as well as $\alpha=1/\veps$ and $\beta=1$ for site $N$. Let us decompose $\hA$ into the contraction of two tensors such that
\begin{equation}\label{eq:psiTauAdecomp}\textstyle
	\bra n,\tilde{n}|\hA|n',\tilde{n}'\ket
	\stackrel{\text{SVD}}{=}
	\sum_{k=1}^m s_k \bra n|\hU|n',k\ket \bra \tilde{n}|\hV|k,\tilde{n}'\ket,\quad\text{where}\quad
	n,\tilde{n},n',\tilde{n}'\in[1,m].
\end{equation}
The tensors $\hU$ and $\hV$ both have bond dimension $m$ and site Hilbert-space dimension $m$, whereas $\hA$ has site dimension $m^2$. Let $\{\hD_k\}$ be a set of $m$ operators on $\V$ that, in the basis $\{|n\ket\}$, are real and diagonal with orthonormality $\Tr(\hD_k\hD_{k'})=\delta_{k,k'}$, and with $\hD_1=\id/\sqrt{m}$ as the first element. Then, a valid choice for the decomposition \eqref{eq:psiTauAdecomp} is
\begin{subequations}\label{eq:psiTauAdecomp2}
\begin{align}
	&\bra n|\hU|n',k\ket=\bra n|\hD_k |n'\ket, \ \
	\bra n|\hV|k,n'\ket=\bra n|\hD_k |n'\ket,\quad
	s_1=m\beta+\alpha-\beta, \ \ s_{k>1}=\alpha-\beta\\
	&\textstyle\text{as}\quad
	\hD_1\otimes\hD_1=\id/m \quad \text{and}\quad
	\sum_{k=1}^m \hD_k\otimes\hD_k=\sum_{n=1}^m|n,n\ket\bra n,n|.
\end{align}
\end{subequations}
\begin{figure*}[t]
\label{fig:psiTauDecomp}
\includegraphics[width=0.97\textwidth]{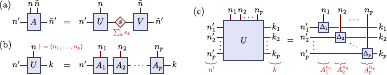}
\caption{Decomposition of the tensors $A^{n,\tilde{n}}$ with $n,\tilde{n}\in[1,m]$ [Eq.~\eqref{eq:psiTauA}] for the MPS $|\psi_\tau\ket$ into a contraction of tensors $A^{n_i}_i$ with smaller site Hilbert spaces, where $n_i\in[1,d_i]$. (a) In an SVD, $A^{n,\tilde{n}}$ is brought to the form $A^{n,\tilde{n}}=U^{n}\operatorname{diag}(s_1,\dotsc,s_m)V^{\tilde{n}}$; cf.\ Eq.~\eqref{eq:psiTauAdecomp}. (b,c) In a second step, $U$ and $V$ are decomposed further into contractions of tensors $A_i$ according to Eq.~\eqref{eq:psiTauUdecomp} by imposing suitable tensor product structures for the bond spaces and site Hilbert spaces.}
\end{figure*}

\textbf{(b)} We can now decompose the tensors $\hU$ and $\hV$ further to obtain an MPS with site dimensions smaller than $m$. To this purpose choose a prime factorization $m=d_1d_2\dotsb d_p$ with any ordering of the factors, and consider a corresponding tensor-product structure $\V=\V_1\otimes\V_2\dotsb\otimes\V_p$ of the space $\V$ with $\dim\V_i=d_i$, corresponding orthonormal bases $\{|n_i\ket\,|\,n_i=1,\dotsc,d_i\}$, and orthonormal bases of diagonal real operators $\{\hD^{(i)}_{k_i}\,|\,k_i=1,\dotsc,d_i\}$, where $\hD^{(i)}_{1}=\id/\sqrt{d_i}$. With the multi-index $k=(k_1,\dotsc,k_p)$, we can choose $\hD_k=\hD^{(1)}_{k_1}\otimes\dotsb\otimes \hD^{(p)}_{k_{p}}$. With multi-indices $n=(n_1,\dotsc,n_p)$ and $n'=(n'_1,\dotsc,n'_p)$, we obtain a decomposition of $\hU$ into a contraction of $p$ MPS tensors (a matrix product),
\begin{subequations}\label{eq:psiTauUdecomp}
\begin{align}
	&\bra n|\hU|n',k\ket=\bra n|\hD^{(1)}_{k_1}\otimes\dotsb\otimes \hD^{(p)}_{k_{p}}|n'\ket
	=\left[A^{n_1}_1 A^{n_2}_2 \dotsb A^{n_p}_p\right]_{n',k},\quad\text{where}\\
	&A^{n_i}_i:=\id_1\otimes\dotsb\id_{i-1}\otimes \Delta^{n_i}_i\otimes\id_{i+1}\otimes\dotsb\otimes\id_{p},\quad\text{and}\quad
	[\Delta^{n_i}_i]_{n'_i,k_i}:=\bra n_i|\hD^{(i)}_{k_i}|n'_i\ket
\end{align}
\end{subequations}
with $n_i,n'_i,k_i\in[1,d_i]$.
A graphical representation for this matrix product decomposition is given in Fig.~\ref{fig:psiTauDecomp}. The same works for $\hV$. In this way, we have obtained a fine-grained MPS form for the states $|\psi_\tau\ket$ [Eq.~\eqref{eq:psiTau}] with bond dimension $m$ and prime-number single-site Hilbert space dimensions $d_i$. Blocking the tensors of consecutive sites in the sense that tensors $A_i$ and $A_{i+1}$ are replaced by
$B^{(n_i,n_{i+1})}:=A^{n_i}_iA^{n_{i+1}}_{i+1}$ etc., we can reach MPS representations $|\psi'_\tau\ket$ with arbitrary factorizations $d_{c,1}d_{c,2}\dotsb d_{c,\ell_c}$ of $m^2$ for each cluster as specified in the Proposition.

\textbf{(c)} Finally, for any such fine-grained MPS representation $|\psi'_\tau\ket$ of the state $|\psi_\tau\ket$, we can regain the original MPS representation \eqref{eq:psiTau} by blocking all $\ell_c$ sites of each cluster. With $\lim_{\veps\to 0}|\psi'_\tau\ket$, we obtain a fine-grained form $|\tau'\ket$ of the two-domain state $|\tau\ket$. If $|\tau'\ket$ were in the $\pMPSn$ set for the fine-grained lattice, $|\tau\ket$ [Eq.~\eqref{eq:2DS}] would be in the $\pMPSn$ set for the original (coarser) lattice. The latter is not the case according to Prop.~\ref{prop:pMPSsmall}, which concludes the proof. \qed

\section{Projected entangled-pair states}\label{sec:PEPS}
PEPS \cite{Niggemann1997-104,Nishino2000-575,Martin-Delgado2001-64,Verstraete2004-7,Verstraete2006-96} are a generalization of MPS to $D>1$ spatial dimensions. First, one chooses a connected graph with vertices $i=1,\dots,N$ and the set $\E$ of edges. In contrast to TTNS, the graph generally contains loops. The edges $e\in\E$ are associated with bond vector spaces $\V_e$ of dimensions $m_e$, and each vertex $i$ is associated with a site Hilbert space $\H_i$ of dimension $d_i$. Let us denote orthonormal basis states for $\V_e$ by $|n_e\ket$ with $n_e=1,\dots,m_e$ and orthonormal basis states for $\H_i$ by $|\sigma_i\ket$ with $\sigma_i=1,\dotsc,d_i$.
So the network structure is characterized by $\G=(\E,\{m_e\},\{d_i\})$.

For such a network $\G$, a PEPS $|\Psi\ket$ is defined by assigning tensors $A_i\in\CC^{d_i\times \prod_{e\in\partial i}m_e}$ to all sites $i$, where $\partial i\subset\E$ is the set of edges connected to vertex $i$. The PEPS $|\Psi\ket$ is obtained by contraction of the tensors $A_i$, i.e., summing over all joint bond indices $n_e$. The set of such normalized PEPS for network $\G$ is denoted by $\PEPS(\G)$.
As depicted in Fig.~\ref{fig:PEPSsmall}a, for a square lattice, a common choice is to assign one tensor $A_i$ to each lattice site. Every tensor carries one index $\sigma_i$ for the site Hilbert space basis. In the bulk, the tensors have four bond indices to be contracted with corresponding indices of tensors on the four neighboring sites. For OBC, tensors at the boundary carry fewer bond indices (two or three).

As discussed in Sec.~\ref{sec:pMPSsmall}, one can equivalently construct PEPS as follows \cite{Affleck1988-115,Verstraete2004-7}: For each edge $e\in\partial i$ connected to site $i$, associate a copy $\V_{i,e}\simeq \V_e$ of the corresponding bond vector space to that site. For each edge $e$ connecting sites $i$ and $j$, prepare the tensor product $\V_{i,e}\otimes\V_{j,e}$ in the maximally entangled state $\sum_{n=1}^{m_e} |n, n\ket$. In this way, we obtain the state
\begin{equation}\label{eq:PEPSmuDef}\textstyle
	|\mu\ket:=\bigotimes_{e\in\E}\left(\sum_{n=1}^{m_e} |n, n\ket_e\right).
\end{equation}
See Fig.~\ref{fig:PEPSsmall}a. It is the analog of the state \eqref{eq:pMPSmuDef} that we employed for MPS. Now, let $e_1,\dotsc,e_z$ denote the $z=|\partial i|$ edges connected to site $i$. We can interpret the tensor $A_i$ as a map $\hA_i:\V_{i,e_1}\otimes\dotsb\otimes \V_{i,e_z}\to \H_i$ from the bond vector spaces into the site Hilbert space of site $i$ with $\bra \sigma|\hat{A}_i|n_{1},\dotsc,n_{z}\,\ket=[A_i^{\sigma}]_{n_{1},\dotsc,n_{z}}$. Then, the PEPS can be written as
\begin{equation}\label{eq:PEPSmu}
	|\Psi\ket=\hat{A}_1\otimes\hat{A}_2\otimes\hat{A}_3\otimes\dotsb\otimes\hat{A}_N|\mu\ket,
\end{equation}
as in Eq.~\eqref{eq:pMPSmu} for pMPS.

\subsection{Small bond dimensions \texorpdfstring{$d\geq m^z$}{d>=m**z}}\label{sec:PEPSsmall}
\begin{figure*}[t]
\label{fig:PEPSsmall}
\includegraphics[width=1\textwidth]{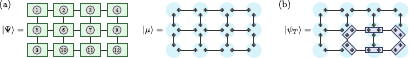}
\caption{(a) A PEPS $|\Psi\ket$ can be constructed by starting from the tensor product $|\mu\ket$ of maximally entangled bond states $\sum_{n=1}^{m_e}|n,n\ket_e$, and applying, subsequently, operators $\hA_i$ on each site that map into the site Hilbert spaces $\H_i$; cf.~Eq.~\eqref{eq:PEPSmuDef}. In the diagrams, physical indices are either indicated by encircled numbers or by encircled dots.
(b) Proposition~\ref{prop:PEPSsmall} establishes the non-closedness of PEPS sets for the case where, at each vertex, the local Hilbert space dimension is equal to or larger than the product of bond dimensions, i.e.,  $d_i\geq m^{z_i}$. It is based on PEPS $|\psi_T\ket$, where the tensors of the pMPS $|\psi_\tau\ket$ [Eq.~\eqref{eq:psiTau}] are applied to $|\mu\ket$ along some loop of the network. Here, we chose the loop $(6,7,8,12,11,10)$.
}
\end{figure*}
By embedding the pMPS constructed in Sec.~\eqref{sec:pMPS} suitably into a PEPS for network $\G$, in generalization of Prop.~\ref{prop:pMPSsmall}, it follows that the sets $\PEPS(\G)$ are generally not closed.
\begin{prop}\label{prop:PEPSsmall}
The sets $\PEPS(\G)$ of normalized PEPS for networks $\G=(\E,\{m_e=m\},\{d_i\})$ are not closed if the network contains at least one loop and if site dimensions obey $d_i\geq m^{z_i}$, where $z_i$ is the degree of vertex $i$.
\end{prop}

\Emph{Proof:}
For the following let $d_i=m^{z_i}$, i.e., $\H_i=\bigotimes_{e\in\partial i}\V_{i,e}$. The generalization to site dimensions $d_i\geq m^{z_i}$ is trivial as before. Let us start from the state $|\mu\ket$ in Eq.~\eqref{eq:PEPSmuDef} for the given network $\G$. It is a PEPS of bond dimension $m$. Acting on $|\mu\ket$ with suitable single-site endomorphisms $\hA_i$, we can define a family of PEPS $|\psi_T\ket$ with unchanged bond and site dimensions. In particular, we choose a closed path (loop) $\mc{C}=(i_1,i_2,\dotsc,i_{L})$ on the graph and apply the endomorphisms $\hA$ and $\hB$ that define the pMPS $|\psi_\tau\ket$ of Eq.~\eqref{eq:psiTau} along that path. This construction of $|\psi_T\ket$ is illustrated in Fig.~\ref{fig:PEPSsmall}b. For example, if the path $\mc{C}$ enters site $i$ from edge $e$ and exits it through edge $e'$, we apply $\hA$ (or $\hB$ for $i=i_{L}$) as defined in Eq.~\eqref{eq:psiTau} to the component $\V_{i,e}\otimes\V_{i,e'}$ of the site-$i$ Hilbert space. With the same considerations as in Sec.~\ref{sec:pMPSsmall}, one finds that the stabilizer $\G_\mu$ of $|\mu\ket$ with respect to the action of invertible operators $\hA_i\in\GL(\H_i)$ has lower dimension than the stabilizer $\G_T$ of $|T\ket:=\lim_{\veps\to 0}|\psi_T\ket$. Here, $|T\ket$ is the two-domain state \eqref{eq:psiTau} on the Hilbert space that corresponds to the loop $\mc{C}$, and consists of maximally entangled states for all other edges. As the single-site reduced density matrices of $|T\ket$ all have full rank, it is a point on the boundary of $\PEPS(\G)$ that is \emph{not} contained in the set, i.e., $\PEPS(\G)$ is not closed. \qed

\subsection{Larger bond dimensions}\label{sec:PEPSlarge}
As discussed for pMPS, also for PEPS simulations, bond dimensions $m_e$ are usually considerably larger than single-site Hilbert space dimensions $d_i$, and we can generalize from Prop.~\ref{prop:PEPSsmall} to this more relevant regime by imposing tensor-product structures for the site Hilbert spaces. The decisive step for pMPS in Sec.~\ref{sec:pMPSlarge}, was to construct a fine-grained version of the state $|\psi_\tau\ket$ [Eq.~\eqref{eq:psiTau}] that yields the two-domain state $|\tau\ket$ in the limit $\veps\to 0$. Suitable tensor decompositions lead to a state $|\psi_\tau'\ket$ for a lattice with smaller site dimensions, allowing for the physically most relevant scenario $d_i\ll m_e$. This fine-grained state $|\psi_\tau'\ket$ is chosen such that a blocking of sites (tensors) recovers $|\psi_\tau\ket$. For PEPS, we need to extend these tensor decompositions of $|\psi_\tau\ket$ one step further.
\begin{figure*}[t]
\label{fig:PEPSlarge}
\includegraphics[width=1\textwidth]{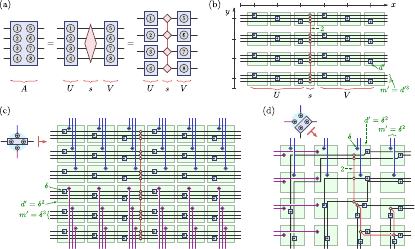}
\caption{(a) For results on PEPS with $d_i< \prod_{e\in\partial i}m_e$, we extend the decomposition of the tensors $A$ in Eq.~\eqref{eq:psiTauA}. As shown in Sec.~\ref{sec:pMPSlarge} and Fig.~\ref{fig:psiTauDecomp}, one can do an SVD $A=UsV$, and decompose the isometries $U$ and $V$ by introducing a suitable tensor product structure for the bond vector spaces $\V=\V_1\otimes\dotsb\otimes\V_p$ and site spaces. Here, we extend this decomposition by bringing the singular-value tensor $s$ into the compatible matrix-product form \eqref{eq:psiTau-s-decomp}. In the shown example, $p=4$.
(b) Such decompositions are employed to obtain fine-grained PEPS versions of the pMPS $|\psi_\tau\ket$, resulting in Prop.~\ref{prop:PEPSbig1} that covers PEPS for cubic lattices in $D$ dimensions with PBC in the $x$ direction. With $p=12$, the diagram shows the decomposition of one tensor $A$ into a tensor network for $2\ell_x=6$ times $L_y=4$ sites. It can be encoded as a block of PEPS tensors as indicated by the green boxes with single-site dimension $d'$ bond dimension $m'=d'^3$ in $x$ direction and arbitrary bond dimensions $\geq 2$ in $y$ direction.
(c,d) Proposition~\ref{prop:PEPSbig2} generalizes to a larger class of PEPS networks using fine-grained versions of states $|\psi_T\ket$ as depicted in Fig.~\ref{fig:PEPSsmall}b. Here we show how the components of $|\psi_T\ket$ are fine-grained.
Panel (c) gives an example for fine-graining site 7 from Fig.~\ref{fig:PEPSsmall}b into a block of $6\times 6$ sites, with new site dimension $d'=\delta^2$ and new bond dimensions $m'=\delta^3$, where the original PEPS $|\psi_T\ket$ has $m=m'^{6}=\delta^{18}$ and $d=d'^{36}=\delta^{72}=m^4$.
Panel (d) gives an example for fine-graining site 6 from Fig.~\ref{fig:PEPSsmall}b into a block of $4\times 4$ sites.
}
\end{figure*}

Before we proceed with these decompositions, let us state clearly what we mean by \emph{blocking} and \emph{fine-graining}: A network $\G'=(\E',\{m'_e\},\{d'_i\})$ is called a fine-grained version of network $\G=(\E,\{m_e\},\{d_i\})$ if there exists a blocking scheme that maps $\G'$ onto $\G$ in the following sense. The graph of $\G'$ is partitioned into clusters, and each cluster $C_i$ with site dimensions $\{d'_{j}\,|\,j\in C_i\}$ corresponds to exactly one vertex $i$ of $\G$ such that $d_i=\prod_{j\in C_i}d'_j$. The edge $e$ in $\G$ that connects vertices $i$ and $j$ corresponds to all edges in $\G'$ that connect clusters $C_i$ and $C_j$. The product of the bond dimensions of the latter must agree with $m_e$. We call $|\Psi'\ket\in\PEPS(\G')$ a fine-grained version of $|\Psi\ket\in\PEPS(\G)$ if $\G'$ is a fine-grained version of $\G$ and if there exists a blocking scheme that maps $|\Psi'\ket$ to $|\Psi\ket$ in the following sense: The PEPS tensor $A_i$ of $|\Psi\ket$ for vertex $i$ is obtained by contracting the PEPS tensors of $|\Psi'\ket$ in cluster $C_i$ and grouping indices of outgoing edges into one multi-index for each connected cluster $C_{j\in\partial i}$. Note that not every $|\Psi\ket\in\PEPS(\G)$ necessarily has a fine-grained pendant in $\PEPS(\G')$, but every $|\Psi'\ket\in\PEPS(\G')$ is a fine-grained version of some state in $\PEPS(\G)$.

In Sec.~\ref{sec:pMPSlarge}, we started from the tensors $\hA=\alpha \hP_d+\beta\hP_o$ that define the pMPS format \eqref{eq:psiTau} of $|\psi_\tau\ket$. With an SVD, we expressed it as a contraction \eqref{eq:psiTauAdecomp} of an isometry $\hU$, the diagonal operator $\hat{s}$ of singular values and a second isometry $\hV$. The isometries carry the physical indices and were decomposed into matrix products \eqref{eq:psiTauAdecomp2} for a fine-grained lattice and a corresponding tensor product structure for the bond vector spaces $\V$. For the following, we want to also decompose $\hat{s}=\sum_{k=1}^m s_k|k\ket\bra k|$ in accordance with the tensor product structure $\V=\V_1\otimes\dotsb\otimes \V_p$ introduced for the decompositions \eqref{eq:psiTauAdecomp2} of $\hU$ and $\hV$. For a bond with $\dim\V=m$, the dimensions $d_i=\dim \V_i$ coincide with the site-dimensions of the fine-grained lattice such that $m=d_1d_2\dotsb d_p$, and we wrote the bond basis as $|k\ket=|k_1,\dotsc,k_p\ket$ with $k_i\in[1,d_i]$. While it is not necessary, we assume here and in the following, for simplicity, that the same factorization of $\V$ is used for both $\hU$ and $\hV$.
The structure of the singular values is very simple, $s_1=m\beta+\alpha-\beta$ and $s_{k>1}=\alpha-\beta$, where $k=1$ corresponds to $k_1=\dotsb =k_p=1$. Hence, $\hat{s}$ can be written in the form of a matrix product operator \cite{Schollwoeck2011-326} with bond dimension 2,
\begin{subequations}\label{eq:psiTau-s-decomp}
\begin{align}
	\hat{s}&\textstyle=s_1|1,1,\dotsc,1\ket\bra 1,1,\dotsc,1|+(s_{k>1}-s_1)\bigotimes_{i=1}^p\id_i\\
	&=\Bmatrix{s_1|1\ket\bra 1|_1 & (s_{k>1}-s_1)\id_1}
	\Bmatrix{|1\ket\bra 1|_2 &0\\ 0& \id_2} \dotsb
	\Bmatrix{|1\ket\bra 1|_{p-1} &0\\ 0& \id_{p-1}}
	\Bmatrix{|1\ket\bra 1|_p \\ \id_p}.
\end{align}
\end{subequations}
The matrix product representation in the second line is analogous to Eqs.~\eqref{eq:psiWA} and \eqref{eq:psiWA-obc}; each entry of the $i$th matrix is an operator on $\V_i$.

In this way, we have obtained a comb-like decomposition of the tensors \eqref{eq:psiTauA} that constitute the pMPS $|\psi_\tau\ket$ as depicted in Fig.~\ref{fig:PEPSlarge}a. The tensors and contraction lines of this fine-grained version can be arranged to form, e.g., a $D$-dimensional cubic lattice as shown in Fig.~\ref{fig:PEPSlarge}b for $D=2$.
\begin{prop}\label{prop:PEPSbig1}
The set $\PEPS(\G)$ of normalized PEPS with a network $\G=(\E,\{m_e\},\{d_i\geq d\})$ for a cubic lattice in $D$ dimensions is not closed under the following sufficient conditions.
In $x$ direction, the network $\G$ has PBC, bond dimension $m_x=d^{\ell_x}$, and length $L_x= 2N_c\ell_x$ with $N_c\geq 3$. For the other $D-1$ spatial directions,
one can choose arbitrary lengths $L_y,L_z,\dotsc\geq 1$, arbitrary boundary conditions, and any bond dimensions $m_y,m_z,\dotsc\geq 2$. 
\end{prop}

\Emph{Proof:}
We can partition the network $\G$ into $N_c$ clusters, defined by cuts on the $x$ axis at $x=2\ell_x$, $4\ell_x$, $\dots$, $2N_c\ell_x$. $\G$ is a fine-graining of the one-dimensional lattice with $N_c$ sites and PBC. For cluster $j\in[1,N_c]$, we can encode a fine-grained version of the $j$th pMPS tensor of $|\psi_\tau\ket$ in PEPS form for the cubic lattice of the cluster with $2\ell_x\times L_y\times L_z\times \dotsb$ sites and bond dimension $m_x=d^{\ell_x}$ in $x$ direction. We only need bond dimension $m_y,m_z,\dotsc\geq 2$ in the other directions, in order to encode the matrix product decomposition \eqref{eq:psiTau-s-decomp} of the singular-value tensor $\hat{s}$. This construction yields a PEPS representation for the fine-grained state $|\psi_\tau'\ket$ and is illustrated in Fig.~\ref{fig:PEPSlarge}b for $D=2$. For any $\veps>0$, $|\psi_\tau'\ket$ is in $\PEPS(\G)$.
Now, blocking all tensors of each cluster into one big tensor, we recover the pMPS representation \eqref{eq:psiTauA} of $|\psi_\tau\ket$. As the two-domain state $\lim_{\veps\to 0}|\psi_\tau\ket$ is not in $\pMPSn(N_c,m'=d^{2\ell_x L_y L_z\dotsb},d'=m'^{2})$, $\lim_{\veps\to 0}|\psi_\tau'\ket$ is not in $\PEPS(\G)$, i.e., $\PEPS(\G)$ is not closed. \qed

This result can be extended to a much bigger class of PEPS networks by fine-graining the states $|\psi_T\ket$ of Sec.~\ref{sec:PEPSsmall} instead of $|\psi_\tau\ket$. The only additional ingredient needed are fine-grained versions of the maximally entangled states $\sum_{n=1}^m|n,n\ket$ that define the state $|\mu\ket$ [Eq.~\eqref{eq:PEPSmuDef}]. Introducing, as before, a tensor product structure $\V=\V_1\otimes\dotsb\otimes \V_p$ with $d_i=\dim\V_i$ and $m=\dim\V=d_1\dotsb d_p$, one can simply use the decomposition
\begin{equation}\textstyle
	\sum_{n=1}^m|n,n\ket=\bigotimes_{i=1}^p\big(\sum_{n_i=1}^{d_i}|n_i,n_i\ket\big).
\end{equation}
\begin{prop}\label{prop:PEPSbig2}
Let $\G$ be any network as characterized in Prop.~\ref{prop:PEPSsmall} and $|\psi_T\ket\in \PEPS(\G)$ a family of PEPS as described in its proof, i.e., states obtained by starting from $|\mu\ket$  and applying the MPS operators \eqref{eq:psiTau} for $|\psi_\tau\ket$ along some loop in $\G$. Let $\G'$ be a fine-grained version of $\G$, consisting of $N_c$ interconnected clusters, each corresponding to one vertex of $\G$. The bond dimensions of $\G'$ need to be able to accommodate the tensors of a fine-grained version $|\psi_T'\ket\in \PEPS(\G')$ of $|\psi_T\ket$. Apart from that, the bond dimensions inside each cluster can be arbitrary. The product of bond dimensions for edges of $\G'$ that get combined into an edge $e$ of $\G$ in the blocking $\G'\to\G$ must be equal to the corresponding bond dimension $m_e$. Under these conditions, the set $\PEPS(\G')$ of normalized PEPS for network $\G'$ is not closed.
\end{prop}
This covers many network structures, for example, (sufficiently large) cubic lattices with open boundary conditions. See Fig.~\ref{fig:PEPSlarge}c,d for exemplary fine-graining schemes.

\section{Tensor network states for infinite system sizes}\label{sec:infiniteTNS}
To directly access the thermodynamic limit, $N\to\infty$, one can use MPS, PEPS and MERA with an infinitely repeated unit cell of tensors (iMPS, iPEPS, iMERA) \cite{Vidal2007-98,Orus2008-78,McCulloch2008_04,Zauner2018-97,Jordan2008-101,Orus2009_05,Evenbly2009-79,Montangero2008-10}. An iMPS with a single-site unit cell, for example, has the same order-3 tensor $A\in \CC^{d\times m\times m}$ for every site of the infinite one-dimensional lattice $\ZZ$, corresponding to the limit $N\to\infty$ in Eq.~\eqref{eq:tiMPS}. An iMPS with a two-site unit cell is characterized by the repetition  $\dots A^{\s_{2i\phantom{+}}}_\text{e}\!\!\! A^{\s_{2i+1}}_\text{o} A^{\s_{2i+2}}_\text{e} A^{\s_{2i+3}}_\text{o} \dots$ of two tensors, one for all even sites and one for all odd sites.

Avoiding more complicated topologies, we consider here the closedness of such sets of TNS $|\Psi\ket$ for infinite systems in the sense that the resulting sets of reduced density matrices $\dm_\A=\Tr_\B|\Psi\ket\bra\Psi|$ are closed or not, where $\A$ is a finite subsystem and $\B$ is its (infinite) complement. This is a relevant topology, since physical observations are for practical reasons limited to finite subsystems.

For an infinite MPS with a transfer operator $E=\sum_\s A^{\s *} \otimes A^\s$ that has only one eigenvalue of magnitude 1, expectation values of local observables and finite-range correlation functions are insensitive to the boundary conditions, since the transfer-matrix product $\lim_{n\to\infty} E^n=|\ell\ket\bra r|$ for and infinite number of sites is completely determined by the left and right eigenvalue-1 eigenvectors of $E$. In this typical scenario, one can hence always imagine working with open boundary conditions on a very large lattice such that the corresponding sets of iMPS density matrices are closed.

Like the sets of pMPS and PEPS, sets of iPEPS are generally not closed due to loops in the networks. In fact, there are more substantial complications: Given the tensor of a translation-invariant iPEPS, it is, in general, undecidable whether the state has a nonzero norm, a certain symmetry, etc.\ \cite{Scarpa2020-125}.

For iMERA, let us distinguish two cases -- iMERA with a finite number of layers (fiMERA) and scale-invariant iMERA (siMERA) where a layer or finite sequence of layers is repeated an infinite number of times:
(a) Sets of reduced density matrices $\dm_\A$ for fiMERA are closed. This is because their $\dm_\A$ can be written as a tensor network that corresponds to the causal cone of $\A$ \cite{Vidal2007-98,Vidal2006}. An example is shown in Fig.~\ref{fig:MERA}. If $\A$ is finite, also the number of tensors in the causal cone is finite, i.e., the $\dm_\A$ are the images of a continuous map from the compact set of the MERA tensors in the causal cone. Hence, they form a closed set.
(b) For siMERA, causal cones of finite subsystems still have finite cross-sections, but they are infinite in the renormalization direction. The density matrix $\dm_\A$ is in this case obtained as the fixed point of the transfer matrix that propagates from layer to layer inside the causal cone \cite{Evenbly2009-79,Montangero2008-10}. As in phase transitions, such fixed points can depend in a non-analytic way on the tensor elements. Hence it is conceivable that the reduced density matrices $\dm_\A$ of siMERA generally form non-closed sets. We leave a detailed analysis for future work.

\section{Discussion and algorithmic remedies for non-closedness}\label{sec:discuss}
\subsection{Summarizing theorem}\label{sec:discussTheorem}
As discussed in the introduction, important objectives of tensor network algorithms are to solve the following optimization problems:
\begin{enumerate}
 \item[(I)] Minimize the distance $\||\Psi\ket-|\psi_0\ket\|^2$ or, equivalently, maximize the overlap $|\bra\psi_0|\Psi\ket|$ for a given state $|\psi_0\ket$ over some set $\M$ of normalized TNS $|\Psi\ket$.
 \item[(II)] Minimize the expectation value $\bra\Psi|\hH|\Psi\ket$ for a given Hamiltonian $\hH$ over some set $\M$ of normalized TNS $|\Psi\ket$.
\end{enumerate}
The findings in Propositions~\ref{prop:oMPS} to \ref{prop:PEPSbig2} can be summarized as follows.
\begin{theorem} Let $\H=\H_1\otimes \dotsb \otimes \H_N$ be the joint Hilbert space of a quantum system with $N$ sites and finite-dimensional site Hilbert spaces $\H_i$.\\[2pt]
\textbf{(a)} Existence: Problems (I) and (II) possess an optimizer for any given $|\psi_0\ket\in\H$ and any given self-adjoint operator $\hH:\H \to \H$ when $\M\subseteq\H$ is any of the closed sets of TNS in Props.~\ref{prop:oMPS} to \ref{prop:MERA} (MPS with OBC, TTNS, or MERA).\\[2pt]
\textbf{(b)} Nonexistence: There exist states $|\psi_0\ket$ and Hamiltonians $\hH$ such that problems (I) and (II) do not possess an optimizer when $\M\subset\H$ is any of the non-closed sets of TNS specified in Props.~\ref{prop:tiMPS} to \ref{prop:PEPSbig2} (translation-invariant MPS with PBC, heterogeneous MPS with PBC, or PEPS) with the corresponding restrictions on the number of sites, site Hilbert space dimensions, and bond dimensions as given in these propositions.
\end{theorem}
\Emph{Proof:}
\textbf{(a)} As was shown, the considered sets of TNS are closed, and they are bounded due to the normalization constraint $\|\Psi\|=1$. Hence, the TNS sets are compact. Now, the assertion follows from the generalized extreme-value theorem, which states that every continuous function on a compact set attains its maximum and minimum in the set \cite{Rudin1976}.
\textbf{(b)} The proofs of Props.~\ref{prop:tiMPS} to \ref{prop:PEPSbig2} establish, in each case, the existence of states $|\psi_0\ket\in\H$ that lie outside the TNS set $\M$ but belong to its closure. For these states, an overlap optimization will approach but never reach $|\psi_0\ket$, i.e., problem (I) has no (global) optimizer. With regard to the energy minimization problem (II), choose $|\psi_0\ket$ as before. Then the expectation-value minimization for the Hamiltonian $\hH=-|\psi_0\ket\bra\psi_0|$ or any other parent Hamiltonian \cite{Fannes1992-144,PerezGarcia2007-7,PerezGarcia2008-8} of $|\psi_0\ket$ reduces to overlap maximization and the assertion follows. \qed

\subsection{Examples with or without optimizers in non-closed TNS sets}
For sets $\M$ of tiMPS, pMPS, and PEPS with loops, we found that the boundary $\partial\M$ generally contains states $|\psi_0\ket$ that are not elements of $\M$. The W state $|W\ket$, discussed in Sec.~\ref{sec:tiMPS}, is an example for tiMPS. The two-domain states $|\tau\ket$ and fine-grained versions of them are examples for pMPS (Sec.~\ref{sec:pMPS}). Embeddings $|T\ket$ of two-domain states into PEPS for higher dimensions and fine-grained versions thereof are examples for PEPS with loops (Sec.~\ref{sec:PEPS}). For these states, maximization of the overlap or minimization of the expectation value for a parent Hamiltonian over the respective TNS set lead to a non-included boundary point.

Both tasks, (I) and (II), of Sec.~\ref{sec:discussTheorem} correspond to the minimization of a convex functional $F(\Psi)$. However, since the considered quasi-projective varieties $\M$ of TNS usually have much lower dimensions than the full Hilbert space $\H$, the optimizers do not need to lie at the boundary $\partial\M$ of the TNS set. Concerning this point, see also Sec.~\ref{sec:geometry} and Fig.~\ref{fig:geometry}.

\footnotetext{T.\ Barthel \emph{et al.}, in preparation.}
The prevalence of the non-existence of optimizers in practical simulation problems is an interesting topic for future research. In the following, let us describe some examples where optimizers are indeed interior points of non-closed TNS sets \cite{Note3}. Specifically, consider the bilinear-biquadratic spin-1 chain
\begin{equation} \label{eq:H_blbq} \textstyle
	\hH_\theta = \sum_{i=1}^N\left[\cos\theta\,\hat{\vec{S}}_i \cdot \hat{\vec{S}}_{i+1} + \sin\theta\, (\hat{\vec{S}}_i \cdot \hat{\vec{S}}_{i+1})^2\right]
\end{equation}
with spin operators $\hat{\vec{S}}_i=(\hS^1_i,\hS^2_i,\hS^3_i)$, $\hat{\vec{S}}_{N+1}\equiv\hat{\vec{S}}_{1}$, $[\hS^a_i,\hS^b_j]=\mathrm{i}\,\delta_{i,j}\sum_{c=1}^3\epsilon_{abc}\hS^c_i$, and $(\hvS_i)^2=2$. 
The competition between the bilinear and biquadratic terms of the Hamiltonian \eqref{eq:H_blbq} leads to a rich groundstate phase diagram, parametrized by the angle $\theta$ \cite{Mikeska2004,Laeuchli2006-74,Binder2020-102}. The model features a gapped dimerized phase ($-3\pi/4 < \theta < -\pi/4$), the gapped Haldane phase ($-\pi/4 < \theta <\pi/4$), an extended critical phase ($\pi/4 < \theta < \pi/2$), and a gapless ferromagnetic ($\pi/2 < \theta < 5\pi/4$). At $\theta=\arctan 1/3\approx 0.102\pi$, the AKLT state $|\Psi_\AKLT\ket$, an unnormalized tiMPS \eqref{eq:tiMPS} with bond dimension $m=2$ and MPS tensor
\begin{equation}\label{eq:AKLT}
	A=\Bmatrix{-|0\ket&\sqrt{2}\,|1\ket\\ -\sqrt{2}\,|{-1}\ket&|0\ket},
\end{equation}
is the exact and unique ground state \cite{Affleck1987-59,Affleck1988-115}, where $|{\sigma=-1,0,1}\ket$ denote the normalized single-site $\hS^3_i$ eigenstates. Except for the AKLT point $\theta=\arctan 1/3$ and except for the ferromagnetic phase, an MPS representation of the ground state requires bond dimension $m>2$. One can show the following about the best $m=2$ MPS groundstate approximation \cite{Note3}:
(a) The AKLT state is a stationary state for the $m=2$ tiMPS \eqref{eq:tiMPS} and pMPS \eqref{eq:pMPS} energy minimization problems for all $\theta$ and any system size $N>1$.
(b) In the thermodynamic limit ($N\to\infty$), the AKLT state is the best $m=2$ tiMPS groundstate approximation for $-\pi/4 <\theta<\pi/4$, i.e., for the entire Haldane phase. For finite $N$, corrections to the boundaries of this $\theta$ interval are exponentially small in $N$.
(c) In the thermodynamic limit, the AKLT state is a minimum of the $m=2$ pMPS single-site alternating least-squares optimization (DMRG) for $\arctan 2-\pi <\theta<\arctan 2\approx 0.35\pi$. For finite $N$, one finds again exponentially small corrections to the interval boundaries.

\subsection{Geometric interpretation and implications for TNS simulations}\label{sec:geometry}
\begin{figure*}[t]
\label{fig:geometry}
\includegraphics[width=0.5\textwidth]{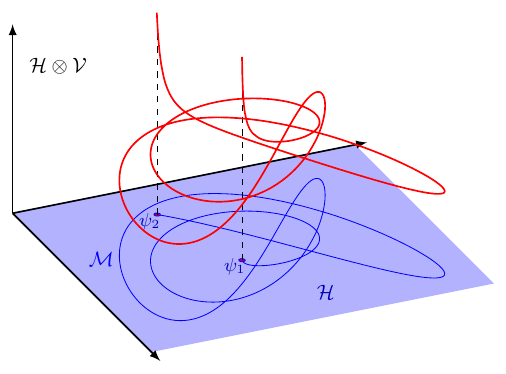}
\caption{The sketch illustrates why sets $\M$ of normalizable TNS can feature exterior boundary points. As an example a pMPS \eqref{eq:pMPS} is determined by the diagonal elements $\Psi^\vs_{n}=[A_1^{\s_1}A_2^{\s_2}\dotsb A_N^{\s_N}]_{n,n}$ with $n=1,\dotsc,m$. They can be interpreted as elements of $\H\otimes\V$, where $\V$ is the $m$ dimensional bond vector space. Diverging elements $\Psi^\vs_{n}$ may cancel when projecting onto the physical Hilbert space $\H$ such that the state $|\Psi\ket=\sum_{n=1}^m \Psi^\vs_n |\vs\ket$ is normalizable. In the illustration, the three-dimensional space represents $\H\otimes\V$, the red curve that diverges to $z\to\infty$ at its ends represents $\Psi^\vs_{n}$, the shaded $xy$ plane represents $\H$, and the blue curve represents the TNS set $\M\subset \H$. Its two boundary points $|\psi_1\ket$ and $|\psi_2\ket$ are not included and are analogs of states like the W state \eqref{eq:Wstate} for tiMPS or the two-domain states \eqref{eq:2DS} for pMPS.}
\end{figure*}
The convergence to an exterior boundary point of a non-closed TNS set is necessarily accompanied by the divergence of tensor elements. This can be concluded by contradiction: If we impose finite bounds on the absolute values of all tensor elements of a TNS, the set of tensors is closed and bounded. The TNS are obtained by doing the tensor contractions. As the latter corresponds to a continuous map, the resulting set of TNS is closed.

Geometrically, exterior boundary points can come about as follows. Consider a pMPS $|\Psi\ket\in\H$ as defined in Eq.~\eqref{eq:pMPS}. Now, $|\Psi\ket$ may have norm one although the tensor $\Psi^\vs_{n}:=[A_1^{\s_1}A_2^{\s_2}\dotsb A_N^{\s_N}]_{n,n}$ has infinite elements. Such infinite elements would have to cancel when taking the trace
\begin{equation}\textstyle
	\bra\vs|\Psi\ket=\Tr\left( \Psi^\vs\right)=\sum_{n=1}^m \Psi^\vs_n
\end{equation}
in Eq.~\eqref{eq:pMPS} that corresponds to the contraction for the bond $(N,1)$. Such cases can lead to exterior boundary points of TNS sets as sketched in Fig.~\ref{fig:geometry}.

Similar scenarios are commonplace for low CP-rank tensor approximations \cite{Hitchcock1927-6,Carroll1970-35,Harshman1970-16} and known, there, under the name CANDECOMP/PARAFAC degeneracies \cite{Paatero2000-14,Stegeman2008-30,Bini1979-8,deSilva2008-30}. In these problems, the non-existence of optimizers is not a rare instance. For given tensor dimensions and desired (reduced) CP-rank, the set of tensors that do not have a best low-rank approximation has positive volume \cite{deSilva2008-30}.

\subsection{Regularizations for optimizations over non-closed TNS sets}
A first idea to resolve the problem of convergence to exterior boundary points might be to drop the normalization constraint $\|\Psi\|=1$ and to impose instead a constraint like $\sum_\s A^\s_i A^{\s\dag}_i=\id$ on the TNS tensors. This would, however, not resolve the issue. When approaching an exterior boundary point of a non-closed TNS set, the norm $\|\Psi\|$ of the state would simply go to zero. 
In order to avoid divergent tensor elements and resulting algorithmic instabilities, we suggest two different regularizations.

Let $F(\Psi)=F(\Psi(\{A_i\}))$ denote either of the functionals $\||\Psi\ket-|\psi_0\ket\|^2$ or $\bra\Psi|\hH|\Psi\ket$ for the optimization problems (I) and (II) of Sec.~\ref{sec:discussTheorem}, respectively. The objective is to minimize $F(\Psi)$ over one of the non-closed sets $\M$ of normalized TNS $|\Psi(\{A_i\})\ket$ specified in Props.~\ref{prop:tiMPS} to \ref{prop:PEPSbig2} ($\tiMPS$, $\pMPSn$, and $\PEPS$). For any normalized state $|\psi_0\ket$ and any norm-bounded Hamiltonian $\hH$, respectively, the corresponding optimization problems for the regularized functional
\begin{equation}\label{eq:regularize1}\textstyle
	\tilde{F}(\Psi(\{A_i\})) = F(\Psi(\{A_i\}))+\sum_i\lambda_i \|A_i\|^2
\end{equation}
have an optimizer in $\M$, where $\lambda_i>0$ are small regularization parameters.
This is because the minimization can be restricted to a sublevel set of $\tilde{F}$ which, due to the boundedness of $F$, is norm-bounded.

One issue with the functional \eqref{eq:regularize1} is that it is not invariant under gauge transformations \eqref{eq:gaugeTrafo} of the TNS tensors. For MPS with PBC, this problem could be alleviated by using instead the product of transfer matrices in the regularization such that
\begin{equation}\label{eq:regularize2}\textstyle
	\tilde{F}(\Psi(\{A_i\})) = F(\Psi(\{A_i\}))+\lambda \|E_1E_2\dotsb E_N\|^2\quad\text{with}\quad
	E_i:=\sum_{\s=1}^d A_i^{\s*}\otimes A_i^{\s}.
\end{equation}
The transfer matrix product $E_1\dotsb E_N$ is invariant with the exception of gauge transformations on bond $(N,1)$. The minimization problems with functional \eqref{eq:regularize2} always have an optimizer in the set $\pMPS(N,m,d)$: An optimization over pMPS \eqref{eq:pMPS} can be seen as an optimization over the $m$ interrelated OBC-MPS
\begin{equation}\textstyle
	|n\ket := \sum_\vs \left[ A_1^{\s_1}A_2^{\s_2}\dotsb A_N^{\s_N}\right]_{n,n}|\vs\ket \quad\text{with}\quad n=1,\dots,m.
\end{equation}
Now, $\bra n|n\ket=[E_1\dotsb E_N]_{(n,n),(n,n)}$ such that the regularization term in Eq.~\eqref{eq:regularize2} enforces a finite norm for each of the OBC-MPS $|n\ket$. According to the discussion of OBC-MPS in Sec.~\ref{sec:oMPS}, this implies that the elements of the tensors $A_i$ cannot diverge.

Another approach was recently suggested in Ref.~\cite{Christandl2021-103}. It uses extensions of TNS such that certain types of boundary points are covered by the extended ansatz. With some computational overhead, one can optimize over theses states using gradient descent or imaginary-time evolution.

\begin{acknowledgments}
We gratefully acknowledge discussions with Matthias Christandl, Daniel S.\ Fran\c{c}a, Hang Huang, Joseph M. Landsberg, Norbert Schuch, and Albert H.\ Werner, the IPAM program \emph{``Tensor methods and emerging applications to the physical and data sciences''}, which initiated this project, as well as support through US Department of Energy grant DE-SC0019449, US National Science Foundation grant DMS-2012286, and US National Science Foundation grant CHE-2037263.\\

This article belongs to the themed \emph{Letters in Mathematical Physics} collection \emph{``Mathematical Physics and Numerical Simulation of Many-Particle Systems''}, V. Bach and L. Delle Site (eds.).
\end{acknowledgments}

\end{document}